\documentclass{article}

\usepackage[utf8]{inputenc}
\usepackage[fleqn]{amsmath}
\usepackage{amsfonts, amssymb}
\usepackage[bookmarks=false,pageanchor=false,hidelinks,pdftex]{hyperref}
 \hypersetup{
     colorlinks,
     linkcolor={red!50!blue},
     citecolor={blue!50!green},
     urlcolor={blue!80!black}
 }
\usepackage[ntheorem]{empheq}
\usepackage[hyperref]{ntheorem}

\newtheorem{theorem}{Theorem}
\newtheorem{lemma}{Lemma}

\newtheorem{corollary}{Corollary}
\newtheorem{definition}{Definition}
\newtheorem{condition}{Condition}

\usepackage{extarrows}                        \usepackage{units} \usepackage{bm} \usepackage{bbold}                    \usepackage{tensor}                     \usepackage{soul} \usepackage{cancel} \usepackage[retainorgcmds]{IEEEtrantools} \usepackage{graphicx}
\usepackage{fixltx2e} 
\usepackage{pbox}
\newcommand{\diagram}[2][{}]{\pbox{\textwidth}{\includegraphics[#1]{{#2}}}}

\usepackage{color}
\usepackage{realboxes}
\usepackage{framed}
\usepackage{array} \usepackage{mathtools} \usepackage[version=3]{mhchem}

\usepackage{enumitem}                                             
\usepackage[numbers,sort&compress]{natbib} \usepackage{bookmark}                       \usepackage{float}

\usepackage{ytableau} 
\usepackage{youngtab}
\usepackage{adjustbox}

\ytableausetup{mathmode, centertableaux}
\usepackage{xargs}
\usepackage{letltxmacro}                          \usepackage{xcolor}
\usepackage{tikz}
\usepackage{tkz-euclide}
\usetikzlibrary{matrix,fit,arrows,decorations}

\usepackage[a4paper,left=2.0cm]{geometry}
\renewenvironment{abstract}{\centering\begin{minipage}{.95\textwidth}
\sffamily{\bf Abstract:}}
{\end{minipage}\vskip 3em}
\makeatletter
\renewcommand\@maketitle{\hfill
\begin{minipage}{\textwidth}
\vskip 2em
{\LARGE \bf \@title \par }
\vskip 1.5em
{\large \@author \par}
\end{minipage}
\vskip 3em \par
}
\makeatother
\usepackage{authblk}

\title{Simplification Rules for Birdtrack Operators}
\author[1]{J. Alcock-Zeilinger}
\author[1]{H. Weigert}
\affil[1]{\small University of Cape Town; Dept. of Physics, Private Bag X3, Rondebosch 7701, South Africa}
\date{October 2016}

\begin{document}
\maketitle

\newcommand{\ud}{\mathrm{d}}
\newcommand{\bra}[1]{\langle #1 \vert}
\newcommand{\ket}[1]{\vert #1 \rangle}
\newcommand{\Tr}[1]{\mathrm{Tr}\left(#1\right)}

\newcommand{\FPic}[2][{}]{\hspace{-0.27mm}\pbox{\textwidth}{\includegraphics[#1]{{#2}}}\hspace{-0.27mm}}

\newcommand{\ybox}[1]{\begin{ytableau} #1 \end{ytableau}}

\newcommand*\circled[1]{\tikz[baseline=(char.base)]{
            \node[shape=circle,draw,inner sep=2pt] (char) {#1};}}

\newcommand{\SUN}{\mathsf{SU}(N)}
\newcommand{\MixedPow}[2]{V^{\otimes
    #1}\otimes\left(V^*\right)^{\otimes #2}}
\newcommand{\Pow}[1]{V^{\otimes #1}}
\newcommand{\DAlg}[1]{\left(V^*\right)^{\otimes #1}}
\newcommand{\Lin}[1]{\mathrm{Lin}\left( #1 \right)}
\newcommand{\API}[1]{\mathsf{API}\left( #1 \right)}
\newcommand{\InvAlg}[1]{A\left[ S_{#1} \right]}
\newcommand{\Rsim}{\stackrel{\mathcal{R}}{\sim}}

\newcommand{\Ampc}[1]{\cancel{#1}_c\left[\mathcal{R}\right]}
\newcommand{\Ampr}[1]{\cancel{#1}_r\left[\mathcal{C}\right]}
\newcommand{\qed}{\hfill\tikz{\draw[draw=black,line width=0.6pt] (0,0) rectangle (2.8mm,2.8mm);}\bigskip}

\def\smath#1{\text{\scalebox{.8}{$#1$}}}
\def\sfrac#1#2{\smath{\frac{#1}{#2}}}

\let\oldytableau\ytableau
\let\endoldytableau\endytableau
\renewenvironment{ytableau}{\begin{adjustbox}{scale=.78}\oldytableau}{\endoldytableau\end{adjustbox}}

\makeatletter
\newcommand{\vast}{\bBigg@{3}}
\newcommand{\Vast}{\bBigg@{4}}
\makeatother

\tikzstyle basiclabel=[draw=none,fill=none,shape=rectangle,inner sep=2pt,scale=.8]
\tikzstyle leftlabel=[basiclabel,anchor=east]
\tikzstyle rightlabel=[basiclabel,anchor=west]

\begin{abstract}
  This paper derives a set of easy-to-use tools designed to simplify
  calculations with birdtrack operators comprised of symmetrizers and
  antisymmetrizers.  In particular, we present cancellation rules
  allowing one to shorten the birdtrack expressions of operators, and
  propagation rules identifying the circumstances under which it is
  possible to propagate symmetrizers past antisymmetrizers and vice
  versa. We exhibit the power of these simplification rules by means
  of a short example in which we apply the tools derived in this paper
  on a typical operator that can be encountered in the representation
  theory of $\SUN$ over the product space $\Pow{m}$. These rules form
  the basis for the construction of compact Hermitian Young projection
  operators and their transition operators addressed in companion
  papers~\cite{Alcock-Zeilinger:2016sxc,Alcock-Zeilinger:2016cva}.
\end{abstract}
\vfill

\setlength{\parskip}{2pt plus 2pt minus 1pt}
\tableofcontents
\setlength{\parskip}{7pt plus 2pt minus 1pt}

\pagebreak

\section{Introduction}

In the 1970's Penrose~\cite{Penrose1971Com,Penrose1971Mom} developed a
graphical method of dealing with objects typically encountered in the
representation theory of semi-simple compact Lie groups, as is used in
QFT. This new formalism was
subsequently applied in a collaboration with
MacCallum~\cite{Penrose:1972ia}. It is clear from Penrose's work that
these graphical tools found their inspiration in Feynman diagrams and
thus allow visually intuitive calculations 
of quantities in the QCD context, since $\SUN$ is the gauge group of
QCD.

Penrose's graphical formalism obtained a more modern treatment by
Cvitanovi{\'c}~\cite{Cvitanovic:2008zz} in early part of the $21^{st}$
century. It is Cvitanovi{\'c} who dubbed the diagrams \emph{birdtracks}.

Birdtracks are gaining in their popularity as a computational tool for
a modern treatment of group theory, in particular the representation
theory of semi-simple Lie groups, and their applications to QFT. There
however do not exist any \emph{practical} tools that allow the easy
manipulation of birdtracks in the literature. The authors suspect that
this is the reason why birdtracks are not yet as widely used as they
\emph{ought} to be. This paper aims to narrow this gap by providing
several easy-to-use rules that greatly simplify dealing with birdtrack
operators.

We will lay our focus on operators that are derived from Young
projection operators~\cite{Young:1928}, and the simplification rules
presented in this paper are thus best suited for such operators. The
reason for this is the authors' interest in the applications of these
tools in a QCD context where factorization invariably involves color
singlet projections of Wilson line correlators (see
e.g.~\cite{Marquet:2010cf, Weigert:2003mm, Falcioni:2014pka,
  Bomhof:2006dp} for a varied set of fields with possible
applications). Since $\SUN$ is the gauge group of QCD, Young
projection operators come into play through the theory of invariants,
which relates the irreducible representations of $\SUN$ over $\Pow{n}$
to the Young tableaux of size $n$, see~\cite{Fulton:2004,Tung:1985na}
and other standard textbooks. However, the lack of Hermiticity of
Young projection operators disqualifies them from the application to
QCD calculations~\cite{AlcockZeilinger2016Singlets}.

Keppeler and Sj{\"o}dahl made a first step towards overcoming this
problem in~\cite{Keppeler:2013yla}, where they present an iterative
algorithm to construct Hermitian versions of Young projection
operators in the birdtrack formalism. However, the KS-operators soon
become unwieldy and thus impractical to work with in automated
calculations owing to computing time and memory resources necessary in
their construction and application.

Using the simplification rules presented in this paper, the
KS-operators can be simplified drastially; an example of this is given
in Figure~\ref{fig:MOLDAdvantage}.

This direct application, however, is not where these simplification
rules exhaust their usefulness. Further applications are presented in
a list of companion papers:
\begin{enumerate}
\item In~\cite{Alcock-Zeilinger:2016sxc} we present an alternative
  construction algorithm for Hermitian Young projection operators,
  which directly leads to significantly more compact and explicitly
  Hermitian expressions of the operators.
\item The simplification rules are a crucial prerequisite for an
  algorithm that allows us to construct transition operators between
  (Hermitian) Young projection operators corresponding to equivalent
  irreducible representations of
  $\SUN$~\cite{Alcock-Zeilinger:2016cva} and an orthogonal
  basis for the algebra of invariants on $\Pow{m}$.
\item This orthogonal basis can then be used to form a basis for the
  singlet states necessary to determine all color neutral Wilson line
  correlators~\cite{AlcockZeilinger2016Singlets} which find direct
  applications in many branches of QCD. First applications (in a
  context that can be covered with direct calculations) can be found
  in~\cite{Marquet:2010cf,Lappi:2016gqe}.
\end{enumerate}
In this paper, we present two classes of simplification rules, they
form the foundation for all three companion papers:
\begin{enumerate}
\item rules that determine whether certain symmetrizers or
  antisymmetrizers can be cancelled from an operator
  (section~\ref{sec:CancellationRules}), and
\item rules describing when it is possible to propagate sets of
  (anti-) symmetrizers through certain parts of the operator
  (section~\ref{sec:PropagateRules}).
\end{enumerate}
Each result in these sections is accompanied by an example. In
section~\ref{sec:Conclusion}, Fig.~\ref{fig:MOLDAdvantage}, we exhibit
the applicability of these rules.

Before we set out to describe the simplification rules, we need to lay
the groundwork by summarizing the conventions used in this paper in
the following section.

\section{Notation, conventions and known results}\label{sec:Notation}

There exists a multitude of (sometimes contradicting) nomenclature and
conventions in the literature with regards to Young tableaux,
birdtracks, and related objects. This section serves to clarify the
conventions used in this paper, as well as to collect a list of
previously known results that are needed for this paper.

\subsection{Tableaux}\label{sec:Tableaux}

Consider an arrangement of $m$ boxes filled
with unique integers between $1$ and $k$ (for $k\geq m$) for example,
\begin{equation}
  \begin{ytableau}
    1 & 10 & 3 \\
    \none & \none &  6 & 5 & 4 \\
    \none & \none & \none & 7 \\
    \none & \none & 9 & 2 & 8
  \end{ytableau}
\ .
\end{equation}
In this paper, we will refer to such a construct as a
\emph{semi-standard irregular tableau}.  In particular, the term
``semi-standard'' will refer to the requirement that each number
appears \emph{at most once} within a tableau. A special case of such a
tableau is a \emph{Young tableau}, in which we require $k=m$ and the
boxes to be top-aligned and left-aligned, as well as the numbers in
the boxes to increase within each row from left to right and within
each column from top to bottom,
see~\cite{Tung:1985na,Fulton:1997,Sagan:2000} and many other standard
textbooks.\footnote{In some references, the presently described
  tableau may also be referred to as a \emph{standard} Young tableau,
  for example~\cite{Fulton:1997,Sagan:2000}.} For example
\begin{equation}
  \begin{ytableau}
    1 & 3 & 5 & 6 \\
    2 & 4 & 7 \\
    8
  \end{ytableau}
\end{equation}
is a Young tableau of size $8$. In this paper, we shall denote a Young
tableau by an upper case Greek letter, usually $\Theta$ or $\Phi$, and
a semi-standard irregular tableau by $\Tilde{\Theta}$ or
$\Tilde{\Phi}$. Furthermore, we will denote the set of all Young
tableaux of size $n$ by $\mathcal{Y}_n$.\footnote{The size of the set
  $\mathcal{Y}_{n}$ is finite for \emph{any} integer $n$, as is shown
  in~\cite{Chowla:1951}.} For example,
\begin{equation}
  \mathcal{Y}_3 := \vast\lbrace
  \begin{ytableau}
    1 & 2 & 3
  \end{ytableau}, \quad
  \begin{ytableau}
    1 & 2 \\
    3
  \end{ytableau}, \quad
  \begin{ytableau}
    1 & 3 \\
    2
  \end{ytableau}, \quad
  \begin{ytableau}
    1 \\
    2 \\
    3
  \end{ytableau}
\vast\rbrace
\ .
\end{equation}\ytableausetup{mathmode, boxsize=2em}For a particular Young tableau $\Theta\in\mathcal{Y}_n$, we refer to
$\Theta_{(m)}\in\mathcal{Y}_{n-m}$ (for $m<n$) as the \emph{ancestor tableau}
of $\Theta$ $m$ generataions back if $\Theta_{(m)}$ is obtained from
$\Theta$ by removing the boxes $\ybox{n}$, $\ybox{\scriptstyle n-1}$
\ldots $\ybox{\scriptstyle n-m}$ from $\Theta$. For example, if
\ytableausetup
{mathmode, boxsize=normal}
\begin{equation}
\Theta := 
  \begin{ytableau}
    1 & 2 & 4 & 5 \\
    3 & 6 & 8 \\
    7 & 9
  \end{ytableau} \quad \text{and} \quad \Phi :=
  \begin{ytableau}
    1 & 2 & 4 & 5 \\
    3
  \end{ytableau}
\ ,
\end{equation}
then $\Phi$ is the ancestor tableau of $\Theta$ four generations back
and we write $\Phi=\Theta_{(4)}$.

In this paper, we will need another kind of tableau, namely the
\emph{amputated tableau}, as is described in the following definition:

\begin{definition}[Amputated Tableaux]\label{AmputatedTableaux}
  Let $\tilde{\Theta}$ be a tableau.\footnote{We do not require
    $\tilde{\Theta}$ to be a Young tableau for this definition, a more
    general kind of tableau (e.g. a semi-standard irregular tableau) will
    suffice.} Furthermore, let $\mathcal{R}$ be a particular row in
  $\tilde{\Theta}$ and $\mathcal{C}$ be a particular column in
  $\tilde{\Theta}$. Then, we form the \emph{column-amputated tableau
    of $\tilde{\Theta}$ according to the row $\mathcal{R}$,
    $\cancel{\tilde{\Theta}}_c\left[\mathcal{R}\right]$}, by removing
  all columns of $\tilde{\Theta}$ which do not overlap with the row
  $\mathcal{R}$. Similarly, we form the \emph{row-amputated tableau of
    $\tilde{\Theta}$ according to the column $\mathcal{C}$,
    $\cancel{\tilde{\Theta}}_r\left[\mathcal{C}\right]$}, by removing
  all rows of $\tilde{\Theta}$ which do not overlap with the column
  $\mathcal{C}$.
\end{definition}

It should be noted that if $\Tilde{\Theta}$ is semi-standard, then
$\Ampc{\Tilde{\Theta}}$ and $\Ampr{\Tilde{\Theta}}$ will also be
semi-standard. As an example, consider the semi-standard irregular tableau
\begin{equation}
\label{eq:irreg-tableau-TTheta}
  \tilde{\Theta} = 
  \begin{ytableau}
    *(yellow) 1 & *(yellow) 2 & *(yellow!60!green) 3 & *(yellow) 4 \\
    \none & \none & *(green) 6 & 5 \\
    \none & \none & \none & 7 & 8
  \end{ytableau}
\ ,
\end{equation}
where we have marked the row $\mathcal{R}:=(1,2,3,4)$ in yellow, and
the column $\mathcal{C}:=(3,6)^t$ in green. Then, the column- and
row-amputated tableaux according to $\mathcal{R}$ and $\mathcal{C}$
respectively are given by
\begin{equation*}
  \cancel{\tilde{\Theta}}_c\left[\mathcal{R}\right] = 
  \begin{ytableau}
    *(yellow) 1 & *(yellow) 2 & *(yellow) 3 & *(yellow) 4 \\
    \none & \none &  6 & 5 \\
    \none & \none & \none & 7 
  \end{ytableau}
\ ,
\end{equation*}
where the column $(8)^t$ was removed since it does not have
an overlap with the row $\mathcal{R}=(1,2,3,4)$,
$(1,2,3,4)\cap(8)^t=\emptyset$,\footnote{Where we transferred the
  familiar set-notation to rows of tableaux.} and
\begin{equation*}
  \cancel{\tilde{\Theta}}_r\left[\mathcal{C}\right] =
  \begin{ytableau}
    1 & 2 & *(green) 3 & 4 \\
    \none & \none & *(green) 6 & 5
  \end{ytableau}
\ ,
\end{equation*}
where the row $(7,8)$ was removed from $\tilde{\Theta}$, as it does not have an overlap with
the column $\mathcal{C}=(3,6)^t$, $(3,6)^t\cap(7,8)=\emptyset$.

\subsection{Birdtracks}\label{sec:Birdtracks}

As is clear by the title of this paper, we aim to provide
simplification rules for birdtrack operators. In particular, this
paper focuses on operators comprised of symmetrizers and
antisymmetrizers. In this section, we give a \emph{short} overview of
the birdtrack notation~\cite{Cvitanovic:2008zz} and its correspondence
to Young projection operators~\cite{Tung:1985na}. For a more extensive
introduction to birdtracks, readers are referred
to~\cite{Cvitanovic:2008zz}, which also serves as the main resource for
this section.

For each semi-standard tableau $\tilde{\Theta}$ (be it irregular or Young), one may
construct the corresponding sets of symmetrizers $\mathbf{S}_{\tilde{\Theta}}$ and
antisymmetrizers $\mathbf{A}_{\tilde{\Theta}}$\footnote{If the tableau
  $\tilde{\Theta}$ consists of $m$ boxes filled with unique integers
  between $1$ and $k$ for $k>m$, we will draw an empty index line for
  each integer $\leq k$ not appearing in the tableau $\tilde{\Theta}$
  in birdtrack notation.} -- this is in fact a generalization to the standard
construction principle of symmetrizers and antisymmetrizers
corresponding to Young tableaux~\cite{Cvitanovic:2008zz,Tung:1985na,Young:1928}. Each row $\mathcal{R}$ of the tableau will correspond to a
symmetrizer over the numbers appearing in $\mathcal{R}$, and each
column $\mathcal{C}$ corresponds to an antisymmetrizer over the
numbers in $\mathcal{C}$. For example, the symmetrizer over elements $1$ and $2$,
$\bm{S}_{12}$, corresponds to the tableau 
\begin{equation}
  \begin{ytableau}
    1 & 2
  \end{ytableau}
\ .
\end{equation}
This symmetrizer $\bm{S}_{12}$ is given by
$\frac{1}{2}\left(\mathrm{id}+(12)\right)$, where $\mathrm{id}$ is
the identity and $(12)$ denotes
the transposition swapping elements $1$ and $2$. For example,
$\bm{S}_{12}$ acts on a tensor $T^{ab}$ as 
\begin{equation}
  \bm{S}_{12}T^{ab} = \sfrac{1}{2} \left(T^{ab} + T^{ba}\right).
\end{equation}
Graphically, we 
denote the symmetrizer $\bm{S}_{12}=\frac{1}{2}\left(\mathrm{id}+(12)\right)$ as 
\begin{equation}
  \bm{S}_{12} = \; \sfrac{1}{2} \left(\;\scalebox{0.75}{\FPic{2ArrLeft}\FPic{2IdSN}\FPic{2ArrRight}} + \scalebox{0.75}{\FPic{2ArrLeft}\FPic{2s12SN}\FPic{2ArrRight}}\;\right).
\end{equation}
This operator is read from right to left,\footnote{This is no longer
  strictly true for birdtracks representing primitive invariants of
  $\SUN$ over a mixed product $\MixedPow{m}{n}$, where $V^*$ is the
  dual vector space of $V$. A more informative discussion on this is
  out of the scope of this paper; readers are referred
  to~\cite{Cvitanovic:2008zz}.} as it is viewed to act as a linear
map from the space $V\otimes V$ into itself. In this paper, the
elements of $S_n$ (the permutation group of $n$ objects) and linear
combinations thereof will always be interpreted as elements of
$\mathrm{Lin}\left(V^{\otimes n}\right)$ (the space of linear
maps over $V^{\otimes n}$). Following~\cite{Cvitanovic:2008zz}, we will
refer to the permutations of $S_n$ as the \emph{primitive invariants}
(of $\SUN$ over $\Pow{n}$), and thus denote the real subalgebra of
$\Lin{\Pow{n}}$ that is spanned by these primitive invariants by
$\API{\SUN,\Pow{n}}\subset\Lin{\Pow{n}}$.

Following~\cite{Cvitanovic:2008zz}, we denote a symmetrizer over an
index-set $\mathcal{N}$, $\bm{S}_{\mathcal{N}}$, by an empty (white)
box over the index lines in $\mathcal{N}$. Thus, the symmetrizer
$\bm{S}_{12}$ is denoted by
\scalebox{0.75}{\FPic{2ArrLeft}\FPic{2Sym12SN}\FPic{2ArrRight}}. Similarly,
an antisymmetrizer over an index-set $\mathcal{M}$,
$\bm{A}_{\mathcal{M}}$, is denoted by a filled (black) box over the
appropriate index lines. For example,
\begin{equation}
  \bm{A}_{12}=\scalebox{0.75}{\FPic{2ArrLeft}\FPic{2ASym12SN}\FPic{2ArrRight}}
   \quad \text{corresponds to
    the tableau} \quad
  \begin{ytableau}
    1 \\
    2
  \end{ytableau} \ ,
\end{equation}
since antisymmetrizers correspond to columns of tableaux. It should be
noted that (sets of) (anti-) symmetrizers are Hermitian with respect
to the canonical scalar product on $\Pow{m}$ (inherited from $V$),
that is,
\begin{equation}
  \label{eq:ASymsHC}
  \mathbf{S}_{\Tilde{\Theta}}^{\dagger} = \mathbf{S}_{\Tilde{\Theta}} \quad \text{and}
    \quad \mathbf{A}_{\Tilde{\Theta}}^{\dagger} = \mathbf{A}_{\Tilde{\Theta}}
\ .
\end{equation}
This is easiest seen in the birdtrack formalism where Hermitian
conjugation (with respect to the canonical scalar product) of an
operator $A$ corresponds to flipping $A$ about its vertical axis and
reversing the arrows (followed, in general by complex conjugation, which plays no role in
the real algebra $\API{\SUN,\Pow{m}}$ of interest to us
here)~\cite{Cvitanovic:2008zz}.

For each tableau $\tilde{\Theta}$, one can then define an operator
$\Bar{Y}_{\tilde{\Theta}}$ as the product of
$\mathbf{S}_{\tilde{\Theta}}$ and $\mathbf{A}_{\tilde{\Theta}}$
\begin{equation}
\label{eq:projection-op-general-tableau}
  \Bar{Y}_{\tilde{\Theta}} := 
  \mathbf{S}_{\tilde{\Theta}} \mathbf{A}_{\tilde{\Theta}}
\ ;
\end{equation}
this is in fact the generalization of Young 
operators~\cite{Cvitanovic:2008zz,Tung:1985na,Young:1928} (\emph{c.f.} eq.~\eqref{eq:Young-Def}) to
semi-standard tableaux. As an example, the operator
corresponding to the tableau~\eqref{eq:irreg-tableau-TTheta} is given
by
\begin{equation}
\label{eq:Irreg-Tabl-Ops}
    \tilde{\Theta} = 
  \begin{ytableau}
     1 &  2 & 3 &  4 \\
    \none & \none & 6 & 5 \\
    \none & \none & \none & 7 & 8
  \end{ytableau}
\quad \longrightarrow \quad\Bar{Y}_{\tilde{\Theta}} = 
\scalebox{0.75}{\FPic{8ArrLeft}\FPic{8s56N}\FPic{8Sym1234Sym56Sym78N}\FPic{8s45N}\FPic{8ASym34ASym567N}\FPic{8s456N}\FPic{8ArrRight}}
\ .
\end{equation}
As already alluded to in the previous paragraph, Young projection
operators are merely a special kind of the operators discussed so far,
namely that where $\tilde{\Theta}=\Theta$ is a Young tableau. One
aspect that makes Young projection operators special is that there
exists a unique constant $\alpha_{\Theta}\neq 0$ such
that\footnote{$\alpha_{\Theta}$ is a combinatorial constant involving
  the Hook length of the tableau
  $\Theta$~\cite{Cvitanovic:2008zz,Fulton:1997,Sagan:2000}.}
\begin{equation}
  \label{eq:Young-Def}
  Y_{\Theta} := \alpha_{\Theta} \cdot \underbrace{\mathbf{S}_{\Theta} \mathbf{A}_{\Theta}}_{=\Bar{Y}_{\Theta}}
\end{equation}
is idempotent; the object $Y_{\Theta}$ is referred to as the
\emph{Young projection operator} corresponding to $\Theta$.  For an
operator $\Bar{Y}_{\Tilde{\Theta}}$ corresponding to a semi-standard
irregular tableau $\Tilde{\Theta}$, it is not necessarily true that a
non-zero constant $c$ can be found that would yield
$Y_{\Tilde{\Theta}}:=c\cdot\Bar{Y}_{\Tilde{\Theta}}$
idempotent.\footnote{This is easiest seen by means of an example: It
  can be verified via direct calculation that the operator
  corresponding to \scalebox{0.7}{$
  \begin{ytableau}
    1 & 2 & 3 \\
    \none & \none & 6 & 5 \\
    \none & \none & \none & 4
  \end{ytableau}
  $}
  is not a projection operator.} Therefore, we adapt the following
notation: $\Bar{Y}_{\Tilde{\Theta}}$ shall denote the operator
corresponding to a (semi-standard irregular or Young) tableau
$\Tilde{\Theta}$ according
to~\eqref{eq:projection-op-general-tableau}, while the symbol
$Y_{\Theta}$ will refer to the
\emph{unique} Young projection operator corresponding to $\Theta$ that
is furnished with the appropriate constant $\alpha_{\Theta}$ yielding
$Y_{\Theta}$ to be idempotent, \emph{c.f.} eq.~\eqref{eq:Young-Def}.

Let us now summarize the most important properties of Young projection
operators~\cite{Tung:1985na,Young:1928}:
\begin{enumerate}
\item\label{itm:YoungIdempotency} \emph{Idempotency}: The Young
  projection operator $Y_{\Theta}$ corresponding to a Young tableau $\Theta$ in $\mathcal{Y}_{n}$
  satisfies
\begin{subequations}
\label{eq:Young-Props}
  \begin{equation}
    \label{eq:YoungIdempotency}
Y_{\Theta} \cdot  Y_{\Theta} = Y_{\Theta}
  \hspace{1cm}\text{ for all $n$.} 
  \end{equation}
\item\label{itm:YoungOrthogonality} \emph{Orthogonality}: If $\Theta$
  and $\Phi$ are two Young tableaux in $\mathcal{Y}_{n}$, then the
  corresponding Young projection operators $Y_{\Theta}$ and $Y_{\Phi}$
  are mutually orthogonal as projectors,
  \begin{equation}
    \label{eq:YoungOrthogonality}
    Y_{\Theta} \cdot
  Y_{\Phi} = \delta_{\Theta \Phi} Y_{\Theta} \hspace{1cm} \text{ for } n =1,2,3,4
\ ,
  \end{equation}
 and more generally (for all $n$) if $\Theta$ and $\Phi$ have different shapes.
\item\label{itm:YoungCompleteness} \emph{Completeness}: The Young
  projection operators corresponding to all Young tableaux in
  $\mathcal{Y}_{n}$ sum up do the identity operator on $\Pow{n}$,
  \begin{equation}
    \label{eq:YoungCompleteness}
\sum_{\Theta \in
    \mathcal{Y}_n} P_{\Theta} = \mathbb{1}_n
\ \hspace{1cm} \text{ for  $n =1,2,3,4$ }
  \end{equation}
  but not beyond.
\end{subequations}
\end{enumerate}
Generalizations of the Young projection operators that remove the
restrictions on $n$ on the latter two of these three properties allow
one to fully classify the irreducible representations of $\SUN$ over
$\Pow{n}$ via Young tableaux in
$\mathcal{Y}_n$~\cite{Littlewood:1950,Weyl:1946,Cvitanovic:2008zz,Fulton:2004}.
All these generalization build on the generally valid idempotency
property of Young projectors, which will also be the only property we
will rely on in this paper.

The Hermitian conjugate of a Young projection operator~\eqref{eq:Young-Def}
is given by\footnote{Using the fact that $\alpha_{\Theta}$ is a
  \emph{real} constant,
  see~\cite{Cvitanovic:2008zz,Sagan:2000,Fulton:1997} and other
  standard textbooks.}
\begin{equation}
  \label{eq:YoungOpsHC}
Y_{\Theta}^{\dagger} = \left(\alpha_{\Theta} \cdot \mathbf{S}_{\Theta}
  \; \mathbf{A}_{\Theta}\right)^{\dagger} = \alpha_{\Theta}^{\dagger} \cdot \mathbf{A}_{\Theta}^{\dagger}
  \; \mathbf{S}_{\Theta}^{\dagger} = \alpha_{\Theta} \cdot \mathbf{A}_{\Theta}
  \; \mathbf{S}_{\Theta}
\ .
\end{equation}
In general, sets of symmetrizers and antisymmetrizers corresponding to
a Young tableau do not commute,
\begin{equation}
  \label{eq:ASyms-NonCommute}
\left[ \mathbf{S}_{\Theta}, \mathbf{A}_{\Theta} \right] \neq 0
\end{equation}
implying that Young projection operators are not Hermitian; the lack
of Hermiticity of Young projection operators and the implications thereof is discussed in~\cite{Alcock-Zeilinger:2016sxc}. 

As a last example, we construct the birdtrack Young projection
operator corresponding to the following Young tableau
\begin{equation}
  \label{eq:SymsEx1}
  \Theta =
  \begin{ytableau}
    1 & 3 & 4 \\
    2 & 5
  \end{ytableau}
\ .
\end{equation}
Since $Y_{\Theta}$ must be comprised of symmetrizers corresponding to
the rows of $\Theta$ and antisymmetrizers corresponding to the columns
of $\Theta$, we find that
\begin{equation}
  Y_{\Theta}= \underbrace{2}_{\alpha_{\Theta}} \cdot \bm{S}_{134}\bm{S}_{25}\bm{A}_{12}\bm{A}_{35},
\end{equation}
where $\alpha_{\Theta}=2$ ensures the idempotency of $Y_{\Theta}$. In
birdtrack notation, this Young projection operator becomes
\begin{equation}
  \label{eq:SymsEx2}
  Y_{\Theta} = \underbrace{2}_{\alpha_{\Theta}} \cdot 
\underbrace{
\FPic{5ArrLeft}\FPic{5s234N}\FPic{5Sym123Sym45N}\FPic{5s2453N}\FPic{5ASym12ASym34N}\FPic{5s45N}\FPic{5ArrRight}
}_{\bar{Y}_{\Theta}}
\ ,
\end{equation}
where we have used the bar-notation introduced previously, \emph{c.f.}
eqs.~\eqref{eq:projection-op-general-tableau}
and~\eqref{eq:Young-Def}. The benefit of the bar-notation is that it
allows one to ignore additional scalar factors: Let $O$ be a birdtrack
operator comprised of symmetrizers and antisymmetrizers. Then,
$\Bar{O}$ denotes the graphical part of $O$ only, and we have that
\begin{equation}
  \label{eq:bar-notation}
\omega \cdot \Bar{O} = \Bar{O}
\qquad \text{but} \qquad
 \omega \cdot O \neq O
\end{equation}
for any non-zero scalar $\omega$.

In expression~\eqref{eq:SymsEx2} for
$Y_{\Theta}$ we were able to draw the two symmetrizers underneath each other
since they are \emph{disjoint}, and similarly for the two
antisymmetrizers. In fact, the symmetrizers
(resp. antisymmetrizers) corresponding to a semi-standard tableau will \emph{always}
be disjoint, since each number can occur at most once by the
definition of semi-standard tableaux. 
\ytableausetup{mathmode, boxsize=0.7em}

Any operator $O \in \mathrm{Lin}\left(V^{\otimes n}\right)$
can be embedded into $\mathrm{Lin}\left(V^{\otimes m}\right)$ for $m >
n$ in several ways, simply by letting the embedding act as the
identity on $(m-n)$ of the factors; how to select these factors is a
matter of what one plans to achieve.  The most useful convention for
our purposes is to let $O$ act on the first $n$ factors and
operate with the identity on the remaining last $(m-n)$ factors. We will
call this the \emph{canonical embedding}.  On the level of birdtracks,
this amounts to letting the index lines of $O$ coincide with
the top $n$ index lines of $\mathrm{Lin}\left(V^{\otimes m}\right)$,
and the bottom $(m-n)$ lines of the embedded operator constitute the
identity birdtrack of size $(m-n)$. For example, the operator
$\Bar{Y}_{\begin{ytableau} \scriptstyle 1 & \scriptstyle 2 \\ \scriptstyle
    3 \end{ytableau}}$ is canonically embedded into
$\mathrm{Lin}\left(V^{\otimes 5}\right)$ as
\begin{equation}
  \label{eq:CanonicalEmbedding}
  \FPic{3ArrLeft}\FPic{3Sym12ASym13}\FPic{3ArrRight} 
  \; \hookrightarrow \; 
  \FPic{5ArrLeft}\FPic{5Sym12N}\FPic{5s23N}\FPic{5ASym12N}\FPic{5s23N}\FPic{5ArrRight}
\ .
\end{equation}
Furthermore, we will use the same
symbol $O$ for the operator as well for its embedded
counterpart. Thus, $\Bar{Y}_{\begin{ytableau} \scriptstyle 1 & \scriptstyle
    2 \\ \scriptstyle 3 \end{ytableau}}$ shall denote both the
operator on the left as well as on the right hand side of the
embedding~\eqref{eq:CanonicalEmbedding}.  \ytableausetup {mathmode, boxsize=normal}

Lastly, if a \emph{Hermitian} projection operator $A$ projects onto a
subspace completely contained in the image of a projection
operator $B$, then we denote this as $A\subset B $, transferring the
familiar notation of sets to the associated projection operators. In
particular, $A\subset B$ if and only if
\begin{equation}
  \label{eq:OperatorInclusion1}
  A \cdot B = B \cdot A = A
\end{equation}
for the following reason: If the subspaces obtained by consecutively
applying the operators $A$ and $B$ in any order is the same as that
obtained by merely applying $A$, then the subspaces onto which $A$ and
$B$ project not only need to overlap (as otherwise $A\cdot B=B\cdot
A=0$), but the subspace corresponding to $A$ must be completely
contained in the subspace of $B$ - otherwise the last equality
of~\eqref{eq:OperatorInclusion1} would not hold.

Hermiticity is crucial for these statements: since we have seen that
sets of symmetrizers and anitsymmetrizers individually are
Hermitian,~\eqref{eq:OperatorInclusion1} does hold for such sets: a
symmetrizer $\bm{S}_{\mathcal{N}}$ can be absorbed into a symmetrizer
$\bm{S}_{\mathcal{N'}}$, as long as the index set $\mathcal{N}$ is a
subset of $\mathcal{N'}$, and the same statement holds for
antisymmetrizer~\cite{Cvitanovic:2008zz}. For example,
\begin{equation}
  \FPic{3ArrLeft}\FPic{3Sym12SN}\FPic{3Sym123SN}\FPic{3ArrRight} 
  \; = \;
  \FPic{3ArrLeft}\FPic{3Sym123SN}\FPic{3ArrRight} 
  \; = \; 
  \FPic{3ArrLeft}\FPic{3Sym123SN}\FPic{3Sym12SN}\FPic{3ArrRight}
  \ .
\end{equation}
Thus, by the above notation,
$\bm{S}_{\mathcal{N'}}\subset\bm{S}_{\mathcal{N}}$, if
$\mathcal{N}\subset\mathcal{N'}$. Or, as in our example,
\begin{equation}
\label{eq:ASym-Inclusion}
  \FPic{3ArrLeft}\FPic{3Sym123SN}\FPic{3ArrRight} 
  \; \subset \; 
  \FPic{3ArrLeft}\FPic{3Sym12SN}\FPic{3ArrRight}
  \ .
\end{equation}
In this sense, eq.~\eqref{eq:ASym-Inclusion} is a simplification rule
in its own right, as it allows us to ``cancel'' (anti-) symmetrizers
that can be absorbed into longer (anti-) symmetrizers. In
particular,~\eqref{eq:ASym-Inclusion} implies that the image of any
(anti-) symmetrizer is contained in the image of its ancestor (anti-)
symmetrizers!\footnote{Where we transfer the nomenclature of
  ancestor-tableaux to the corresponding (anti-) symmetrizers.}  This
nested inclusion of ancestor operators breaks down for the
standard Young projection
operators whenever they are not Hermitian~\cite{Alcock-Zeilinger:2016sxc}, as for example
\ytableausetup{mathmode, boxsize=0.7em}
\begin{equation}
  \underbrace{
\frac{4}{3} \cdot \FPic{3ArrLeft}\FPic{3Sym13ASym12}\FPic{3ArrRight}
}_{Y_{
      \begin{ytableau}
        \scriptstyle 1 & \scriptstyle 3 \\
        \scriptstyle 2
      \end{ytableau}
}}
  \cdot 
    \underbrace{
  \FPic{3ArrLeft}\FPic{3ASym12N}\FPic{3ArrRight}
}_{Y_{
      \begin{ytableau}
        \scriptstyle 1 \\
        \scriptstyle 2
      \end{ytableau}
}} 
=
  \frac{4}{3} \cdot
  \FPic{3ArrLeft}\FPic{3Sym13ASym12}\FPic{3ASym12N}\FPic{3ArrRight} 
= 
  \frac{4}{3} \cdot \FPic{3ArrLeft}\FPic{3Sym13ASym12}\FPic{3ArrRight}
\end{equation}
but
\begin{equation}
    \underbrace{
  \FPic{3ArrLeft}\FPic{3ASym12N}\FPic{3ArrRight}
}_{Y_{
      \begin{ytableau}
        \scriptstyle 1 \\
        \scriptstyle 2
      \end{ytableau}
}} 
  \cdot
  \underbrace{
\frac{4}{3} \cdot \FPic{3ArrLeft}\FPic{3Sym13ASym12}\FPic{3ArrRight}
}_{Y_{
      \begin{ytableau}
        \scriptstyle 1 & \scriptstyle 3 \\
        \scriptstyle 2
      \end{ytableau}
}}
=
  \frac{4}{3} \cdot
  \FPic{3ArrLeft}\FPic{3ASym12N}\FPic{3Sym13ASym12}\FPic{3ArrRight} 
  \neq
  \frac{4}{3} \cdot \FPic{3ArrLeft}\FPic{3Sym13ASym12}\FPic{3ArrRight}
  \ ,
\end{equation}
which can be verified by direct calculation. On the other hand,
the image of a Hermitian Young projection operators is contained in
the images of its ancestor Hermitian projectors~\cite{Alcock-Zeilinger:2016sxc}.

The direction of the arrow on the index lines of the birdtrack encode
whether the line acts on the vector space $V$ (arrow pointing from
right to left) or its dual $V^*$ (arrow pointing from left to
right)~\cite{Cvitanovic:2008zz}. In this paper, we will only consider
birdtracks acting on a space $\Pow{m}$ (never on the dual) and
thus only encounter birdtracks with arrows pointing from right to
left. To reduce clutter, we will therefore suppress the arrows
and (for example) simply write
\begin{equation}
  \FPic{3s123SN} \quad \text{when we mean} \quad
  \FPic{3ArrLeft}\FPic{3s123SN}\FPic{3ArrRight} \; .
\end{equation}
\ytableausetup{mathmode, boxsize=normal}

We are now in a position to discuss the main result of this paper: We
describe two classes of simplification rules for birdtrack operators
$O$ comprised of symmetrizers and antisymmetrizers, namely
\begin{enumerate}
\item\label{itm:CancellationRules} \emph{Cancellation rules}: These
  describe a set of rules to \emph{cancel} certain symmetrizers and
  antisymmetrizers within an operator $O$. The usefulness of these
  rules is that they can make a long expression significantly shorter,
  and thus more practical and less computationally expensive to work
  with. These rules are described in
  section~\ref{sec:CancellationRules}.
\item\label{itm:PropagationRules} \emph{Propagation rules}: These
  describe the circumstances under which it is possible to commute a
  particular symmetrizer through a (set of) antisymmetrizer(s), and
  vice versa. These rules can be used to create a situation in which
  the cancellation rules (see part~\ref{itm:CancellationRules}) can be
  used, or to make certain features of a particular operator $O$
  (for example its Hermiticity) explicit. These rules can be found in
  section~\ref{sec:PropagateRules}.
\end{enumerate}

These simplification rules come into their own when they are applied
to birdtrack operators in group-theoretic calculations. For examples,
we extensively used these rules in our papers on a compact construction
of Hermitian Young projection
operators~\cite{Alcock-Zeilinger:2016sxc} and transition
operators~\cite{Alcock-Zeilinger:2016cva}. A further example
is given in section~\ref{sec:Conclusion}
(Fig.~\ref{fig:MOLDAdvantage}).

\section{Cancellation rules}\label{sec:CancellationRules}

\subsection{Cancellation of wedged Young
  projectors}\label{sec:CancelYoungs}

We begin by presenting two main cancellation rules,
Theorem~\ref{thm:CancelWedgedYoung} and Corollary~\ref{thm:Cancel-Ops}. The benefit of these rules is that they
can be used to shorten the birdtrack-expressions of certain operators
(sometimes inducing a constant factor), and thus make the resulting
expression more useful for practical calculations.

\begin{theorem}[cancellation of wedged Young projectors]
\label{thm:CancelWedgedYoung}
Consider an operator $O$ consisting of an alternating product of
altogether four symmetrizers and anti-sym\-metrizers, with the
middle pair being proportional to a Young projection operator
  \begin{equation}
    \label{eq:CancelWedgedYoung}
    O = \; \mathbf{A}_{\Phi_1} \; \mathbf{S}_{\Theta} \;
    \mathbf{A}_{\Theta} \; \mathbf{S}_{\Phi_2}
    = \; \mathbf{A}_{\Phi_1} \; \Bar Y_{\Theta} \; \mathbf{S}_{\Phi_2}
  \end{equation}
  such that $\mathbf{S}_{\Theta}\supset\mathbf{S}_{\Phi_2}$ and
  $\mathbf{A}_{\Theta}\supset\mathbf{A}_{\Phi_1}$ i.e. $\mathbf
  S_\Theta \mathbf S_{\Phi_2} = \mathbf S_{\Phi_2} = \mathbf
  S_{\Phi_2} \mathbf S_\Theta$ and $\mathbf A_\Theta \mathbf
  A_{\Phi_1} = \mathbf A_{\Phi_1} = \mathbf A_{\Phi_1} \mathbf
  A_\Theta$ (c.f. eq.~\eqref{eq:OperatorInclusion1}). Then, we can
  drop $\Bar{Y}_\Theta$ while acquiring a scalar factor $1/\alpha_\Theta$:
\begin{equation}
  \label{eq:wedged-Young}
  \mathbf{A}_{\Phi_1} \; \Bar Y_{\Theta} \; \mathbf{S}_{\Phi_2} 
  = 
  \frac1{\alpha_\Theta}
   \mathbf{A}_{\Phi_1} \;  \mathbf{S}_{\Phi_2}
   \ .
\end{equation}
Corresponding cancellations apply if all symmetrizers are exchanged
for antisymmetrizers and vice versa.

Using $Y_\Theta$ instead $\Bar Y_\Theta$ removes the constant. The
form presented here is that usually encountered in practical
calculations.
\end{theorem}
Before looking at a general proof for this statement, we will develop
the strategy for it through an example. To this end take $O$ to be
\begin{equation}
  \label{eq:ExWedgedAnc1}
 O = \;
 \scalebox{0.75}{$\underbrace{\FPic{5s23N}\FPic{5ASym12ASym34N}\FPic{5s23N}}_{\mbox{\normalsize
   $\mathbf{A}_{\Phi_1}$}}\underbrace{\FPic{5Sym12SN}}_{\mbox{\normalsize
   $\mathbf{S}_{\Theta}$}}\underbrace{\FPic{5s23N}\FPic{5ASym12N}\FPic{5s23N}}_{\mbox{\normalsize
 $\mathbf{A}_{\Theta}$}}\underbrace{\FPic{5s354N}\FPic{5Sym123N}\FPic{5s345N}}_{\mbox{\normalsize
$\mathbf{S}_{\Phi_2}$}}$}
\ .
\end{equation}
The central sets of symmetrizers and antisymmetrizers correspond to
the Young tableau
\begin{equation}
  \label{eq:ExWedgedAnc2}
  \Theta =
\begin{ytableau}
 1 & 2 \\
 3
\end{ytableau}
\ ,
\end{equation}
embedded into $\Lin{\Pow{5}}$.
The inclusion criterion can be verified in multiple ways:
\begin{itemize}
\item Thinking in terms of \emph{image} inclusions we note that
$\mathbf{S}_{\Theta}\supset\mathbf{S}_{\Phi_2}$ (since $\mathbf{S}_{\Theta}=\lbrace\bm{S}_{12}\rbrace\supset\lbrace\bm{S}_{125}\rbrace=\mathbf{S}_{\Phi_2}$) and

and
  $\mathbf{A}_{\Theta}\supset\mathbf{A}_{\Phi_1}$ (since $\mathbf{A}_{\Theta}=\lbrace\bm{A}_{13}\rbrace\supset\lbrace\bm{A}_{13},\bm{A}_{24}\rbrace=\mathbf{A}_{\Phi_1}$)
\item Equivalently, in terms of birdtracks we see that
\begin{equation}
  \label{eq:Wedge-Example-Conditions}
 \scalebox{0.75}{$
\underbrace{\FPic{5Sym12R}}_{\mbox{\normalsize
   $\mathbf{S}_{\Theta}$}}\underbrace{\FPic{5s354N}\FPic{5Sym123N}\FPic{5s345N}}_{\mbox{\normalsize
$\mathbf{S}_{\Phi_2}$}}$}
\; = \; 
\scalebox{0.75}{$
\underbrace{\FPic{5s354N}\FPic{5Sym123N}\FPic{5s345N}}_{\mbox{\normalsize
$\mathbf{S}_{\Phi_2}$}}$}
\; = \; 
\scalebox{0.75}{$
\underbrace{\FPic{5s354N}\FPic{5Sym123N}\FPic{5s345N}}_{\mbox{\normalsize
$\mathbf{S}_{\Phi_2}$}}
\underbrace{\FPic{5Sym12R}}_{\mbox{\normalsize
   $\mathbf{S}_{\Theta}$}}$}
\hspace{1cm} \text{and} \hspace{1cm}
 \scalebox{0.75}{$
\underbrace{\FPic{5s23N}\FPic{5ASym12ASym34N}\FPic{5s23N}}_{\mbox{\normalsize
   $\mathbf{A}_{\Phi_1}$}}
\underbrace{\FPic{5s23N}\FPic{5ASym12N}\FPic{5s23N}}_{\mbox{\normalsize
 $\mathbf{A}_{\Theta}$}}$}
\; = \; 
\scalebox{0.75}{$\underbrace{\FPic{5s23N}\FPic{5ASym12ASym34N}\FPic{5s23N}}_{\mbox{\normalsize
   $\mathbf{A}_{\Phi_1}$}}$}
\; = \;
\scalebox{0.75}{$\underbrace{\FPic{5s23N}\FPic{5ASym12N}\FPic{5s23N}}_{\mbox{\normalsize
 $\mathbf{A}_{\Theta}$}}
\underbrace{\FPic{5s23N}\FPic{5ASym12ASym34N}\FPic{5s23N}}_{\mbox{\normalsize
   $\mathbf{A}_{\Phi_1}$}}$}
\ .
\end{equation}
\end{itemize}
Let us explore how the cancellation of eq.~\eqref{eq:wedged-Young}
comes about in example~\eqref{eq:ExWedgedAnc1}: First note that due to
eq.~\eqref{eq:Wedge-Example-Conditions} we may rewrite $O$ as
\begin{equation}
  \begin{tikzpicture}[baseline=(current bounding box.west),
  every node/.style={inner sep=0pt,outer sep=-1pt}        ]
    \matrix(ID)[
    matrix of math nodes,
    ampersand replacement=\&,
    row sep =0mm,
    column sep =0mm
    ]
    { O = \; \underbrace{\scalebox{0.75}{\FPic{5s23N}\FPic{5ASym12ASym34N}\FPic{5s23N}
}}_{\mathbf{A}_{\Phi_1}}
      \& \underbrace{\scalebox{0.75}{\FPic{5Sym12R}}}_{\mathbf{S}_{\Theta}}
      \& \underbrace{\scalebox{0.75}{\FPic{5s23N}\FPic{5ASym12N}\FPic{5s23N}}}_{\mathbf{A}_{\Theta}}
      \&
      \underbrace{\scalebox{0.75}{\FPic{5s354N}\FPic{5Sym123N}\FPic{5s345N}}}_{\mathbf{S}_{\Phi_2}}
      \& \; \xlongequal[\eqref{eq:Wedge-Example-Conditions}]{\text{eq.}} \; 
      \& \scalebox{0.75}{\FPic{5s23N}\FPic{5ASym12ASym34N}\FPic{5s23N}}
      \& {\colorbox{blue!25}{\scalebox{0.75}{\FPic{5s23N}\FPic{5ASym12N}\FPic{5s23N}}}}
      \& \scalebox{0.75}{\FPic{5Sym12N}}
      \& \scalebox{0.75}{\FPic{5s23N}\FPic{5ASym12N}\FPic{5s23N}}
      \& {\colorbox{blue!25}{\scalebox{0.75}{\FPic{5Sym12N}}}}
      \& \scalebox{0.75}{\FPic{5s354N}\FPic{5Sym123N}\FPic{5s345N}}
      \& \; = \; 
      \& \scalebox{0.75}{\FPic{5s23N}\FPic{5ASym12ASym34N}\FPic{5s23N}}
            \underbrace{\left( \scalebox{0.75}{\FPic{5s23N}\FPic{5ASym12N}\FPic{5s23N}\FPic{5Sym12N}}
  \right)}_{\bar{Y}_{\Theta}^{\dagger}} 
            \underbrace{\left( \scalebox{0.75}{\FPic{5s23N}\FPic{5ASym12N}\FPic{5s23N}\FPic{5Sym12N}}
  \right)}_{\bar{Y}_{\Theta}^{\dagger}} 
            \scalebox{0.75}{\FPic{5s354N}\FPic{5Sym123N}\FPic{5s345N}}
            \ .
      \\
};
\draw[-stealth,thick,red] (ID-1-6.north) |- ++(0,+3.5mm)   -|
(ID-1-7.north) node[pos=.2,anchor=south,yshift=1mm]
{\scriptsize$\mathbf{A}_{\Phi_1}\to
  \mathbf{A}_{\Phi_1}\mathbf{A}_{\Theta}$};
\draw[stealth-,thick,red] (ID-1-10.north) |- ++(0,+2.7mm)   -|
(ID-1-11.north) node[pos=.2,anchor=south,yshift=1mm]
{\scriptsize$\mathbf{S}_{\Phi_2}\to\mathbf{S}_{\Theta}\mathbf{S}_{\Phi_2}$};
\end{tikzpicture} 
\end{equation}
Idempotency of $Y_\Theta$ implies
$\bar{Y}_{\Theta}^{\dagger}\bar{Y}_{\Theta}^{\dagger}=\nicefrac{1}{\alpha_{\Theta}}\bar{Y}_{\Theta}^{\dagger}$
so that
\begin{equation}
  \begin{tikzpicture}[baseline=(current bounding box.west),
  every node/.style={inner sep=0pt,outer sep=-1pt}        ]
    \matrix(ID)[
    matrix of math nodes,
    ampersand replacement=\&,
    row sep =0mm,
    column sep =0mm
    ]
    { O = \frac{1}{\alpha_{\Theta}} \cdot \;
      \& \scalebox{0.75}{\FPic{5s23N}\FPic{5ASym12ASym34N}\FPic{5s23N}}
      \& \scalebox{0.75}{\FPic{5s23N}\FPic{5ASym12N}\FPic{5s23N}}
      \& \scalebox{0.75}{\FPic{5Sym12N}}
      \& \scalebox{0.75}{\FPic{5s354N}\FPic{5Sym123N}\FPic{5s345N}}
      \& \; \;
      \xlongequal[\eqref{eq:Wedge-Example-Conditions}]{\text{eq.}} \; \;
      \& \frac{1}{\alpha_{\Theta}} \cdot \underbrace{\scalebox{0.75}{\FPic{5s23N}\FPic{5ASym12ASym34N}\FPic{5s23N}}}_{\mathbf{A}_{\Phi_1}}
      \&
      \underbrace{\scalebox{0.75}{\FPic{5s354N}\FPic{5Sym123N}\FPic{5s345N}}}_{\mathbf{S}_{\Phi_2}}
      \ .
      \\
};
\draw[stealth-,thick,red] (ID-1-2.north) |- ++(0,+3mm)   -|
(ID-1-3.north) node[pos=0.2,yshift=2mm]
{\scriptsize$\mathbf{A}_{\Phi_1}\mathbf{A}_{\Theta}\to
  \mathbf{A}_{\Phi_1}$};
\draw[-stealth,thick,red] (ID-1-4.south) |- ++(0,-3mm)   -|
(ID-1-5.south) node[pos=1,anchor=north,yshift=-4mm]
{\scriptsize\;$\mathbf{S}_{\Theta}\mathbf{S}_{\Phi_2}\to\mathbf{S}_{\Phi_2}$};
\end{tikzpicture} 
\end{equation}

The calculation exhibits a clear three step pattern that immediately furnishes the general proof:
\begin{enumerate}
\item Factor $\mathbf S_\Theta$ from $\mathbf S_{\Phi_2}$ and $\mathbf
  A_\Theta$ from $\mathbf A_{\Phi_1}$ to generate
  $\bar{Y}_{\Theta}^{\dagger}\bar{Y}_{\Theta}^{\dagger}$ (this is
  possible since $\mathbf{S}_{\Theta}\supset\mathbf{S}_{\Phi_2}$ and
  $\mathbf{A}_{\Theta}\supset\mathbf{A}_{\Phi_1}$ as required by the Theorem)
 \begin{equation}
\begin{tikzpicture}[baseline=(current bounding box.west),
  every node/.style={inner sep=1pt,outer sep=-1pt}        ]
    \matrix(ID)[
    matrix of math nodes,
    ampersand replacement=\&,
    row sep =0mm,
    column sep =0mm
    ]
    { O = 
      \& \mathbf{A}_{\Phi_1}
      \& \mathbf{A}_{\Theta}
      \& \mathbf{S}_{\Theta}
      \& 
      \& \mathbf{A}_{\Theta} 
      \& \mathbf{S}_{\Theta}
      \& \mathbf{S}_{\Phi_2}
      \ ,
      \\
};
\draw[-stealth,thick,red] (ID-1-2.north) |- ++(0,+3mm)   -|
(ID-1-3.north) node[pos=.2,anchor=south,yshift=1mm]
{\scriptsize$\mathbf{A}_{\Phi_1}\to
  \mathbf{A}_{\Phi_1}\mathbf{A}_{\Theta}$};
\draw[stealth-,thick,red] (ID-1-7.north) |- ++(0,+3mm)   -|
(ID-1-8.north) node[pos=.2,anchor=south,yshift=1mm]
{\scriptsize$\mathbf{S}_{\Phi_2}\to \mathbf{S}_{\Theta}\mathbf{S}_{\Phi_2}$};
\draw[decorate,decoration={brace,amplitude=4pt},thick] (ID-1-4.south east) --
(ID-1-3.south west) node[pos=.5,anchor=north,yshift=-2mm]
{\scriptsize$\bar{Y}_{\Theta}^{\dagger}$};
\draw[decorate,decoration={brace,amplitude=4pt},thick] (ID-1-7.south east) --
(ID-1-6.south west) node[pos=.5,anchor=north,yshift=-2mm]
{\scriptsize$\bar{Y}_{\Theta}^{\dagger}$};
\end{tikzpicture}
 \end{equation}
\item use idempotency of $Y_\Theta$ so simplify  $\bar{Y}_{\Theta}^{\dagger}\bar{Y}_{\Theta}^{\dagger}=\nicefrac{1}{\alpha_{\Theta}}\bar{Y}_{\Theta}^{\dagger}$
\begin{equation}
\begin{tikzpicture}[baseline=(current bounding box.west),
  every node/.style={inner sep=1pt,outer sep=-1pt}        ]
    \matrix(ID)[
    matrix of math nodes,
    ampersand replacement=\&,
    row sep =0mm,
    column sep =0mm
    ]
    { O = \frac{1}{\alpha_{\Theta}} \cdot
      \& \mathbf{A}_{\Phi_1}
      \& \mathbf{A}_{\Theta}
      \& \mathbf{S}_{\Theta}
      \& \mathbf{S}_{\Phi_2}
      \ ,
      \\
};
\draw[decorate,decoration={brace,amplitude=4pt},thick] (ID-1-4.south east) --
(ID-1-3.south west) node[pos=.5,anchor=north,yshift=-2mm]
{\scriptsize$\bar{Y}_{\Theta}^{\dagger}$};
\end{tikzpicture}
\end{equation}
\item reabsorb  $\mathbf S_\Theta$ into $\mathbf S_{\Phi_2}$ and $\mathbf A_\Theta$ into $\mathbf A_{\Phi_1}$
\begin{equation}
\begin{tikzpicture}[baseline=(current bounding box.west),
  every node/.style={inner sep=1pt,outer sep=-1pt}        ]
    \matrix(ID)[
    matrix of math nodes,
    ampersand replacement=\&,
    row sep =0mm,
    column sep =0mm
    ]
    { O = \frac{1}{\alpha_{\Theta}} \cdot
      \& \mathbf{A}_{\Phi_1}
      \& \; \cancel{\mathbf{A}_{\Theta}}
      \& \; \cancel{\mathbf{S}_{\Theta}}
      \& \; \mathbf{S}_{\Phi_2}
      \& = \frac{1}{\alpha_{\Theta}} \cdot
      \& \mathbf{A}_{\Phi_1} 
      \& \mathbf{S}_{\Phi_2}
      \ .
      \\
};
\draw[stealth-,thick,red] (ID-1-2.north) |- ++(0,+3mm)   -|
(ID-1-3.north) node[pos=.2,anchor=south,yshift=1mm]
{\scriptsize$\mathbf{A}_{\Phi_1}\mathbf{A}_{\Theta}\to
  \mathbf{A}_{\Phi_1}$};
\draw[-stealth,thick,red] (ID-1-4.south) |- ++(0,-3mm)   -|
(ID-1-5.south) node[pos=0.3,anchor=north,yshift=-1mm]
{\scriptsize$\mathbf{S}_{\Theta}\mathbf{S}_{\Phi_2}\to\mathbf{S}_{\Phi_2}$};
\end{tikzpicture}
\end{equation}
\end{enumerate}
\qed

In some applications one finds the ingredients of
Theorem~\ref{thm:CancelWedgedYoung} embedded into chains of Young
projectors~\cite{Keppeler:2013yla,Alcock-Zeilinger:2016sxc},
we thus explicitly formulate the following Corollary:

\begin{corollary}[cancellation of wedged ancestor-operators]\label{CancelWedgedParentOp}
  Consider two Young tableaux $\Theta$ and $\Phi$ such that they have
  a common ancestor tableau $\Gamma$. Let $Y_{\Theta}$, $Y_{\Phi}$
  and $Y_{\Gamma}$ be their respective Young projection operators, all
  embedded in an algebra that encompasses all three. Then
    \begin{equation}
      \label{eq:CancelWedgedParentOp}
Y_{\Theta} Y_{\Gamma} Y_{\Phi} = Y_{\Theta} Y_{\Phi}
\ .
    \end{equation}
  \end{corollary}

  \noindent This Corollary immediately follows from
  Theorem~\ref{thm:CancelWedgedYoung} since the product
  $Y_{\Theta}Y_{\Gamma}Y_{\Phi}$ will be of the form
\begin{equation}
  Y_{\Theta} Y_{\Gamma} Y_{\Phi} =
  \alpha_{\Theta}\alpha_{\Gamma}\alpha_{\Phi} \cdot
  \mathbf{S}_{\Theta}
  \underbrace{{\colorbox{red!20}{$\mathbf{A}_{\Theta} \; \mathbf{S}_{\Gamma}
  \; \mathbf{A}_{\Gamma} \; \mathbf{S}_{\Phi}$}}}_{O} \mathbf{A}_{\Phi}
\ ,
\end{equation}
where the marked factor constitutes $O$ as defined in
equation~\eqref{eq:CancelWedgedYoung} in Theorem~\ref{thm:CancelWedgedYoung}.

\subsection{Cancellation of factors between bracketing sets}

In this section, we present another cancellation Theorem that allows
us to significantly shorten certain operators. The results presented
here follow immediately from a result given in~\cite[Lemma IV.5]{Tung:1985na},
which we paraphrase here in Lemma~\ref{TungLemmaIV.5}. Before we can
give~\cite{Tung:1985na}'s result, we need to define
\emph{horizontal} and \emph{vertical} permutations of a Young tableau:

\begin{definition}[horizontal and vertical
  permutations]\label{HVPermutation}
  Let $\Tilde{\Theta}$ be a semi-standard (Young or irregular) tableau
  such that $n$ is the largest integer appearing in $\Tilde{\Theta}$. Then,
  $\mathbf{h}_{\Tilde{\Theta}}$ shall denote the subset of all permutations in
  $S_n$ that only operate within the rows of $\Tilde{\Theta}$; i.e. that do
  not swap numbers across rows.  We call this the set the
  \emph{horizontal permutations} of $\Tilde{\Theta}$. Similarly, we define the
  set of \emph{vertical permutations} of $\Tilde{\Theta}$,
  $\mathbf{v}_{\Tilde{\Theta}}$, to be the subset of permutations in $S_n$
  that only operate within the columns of $\Tilde{\Theta}$, i.e. those that do
  not swap numbers across columns.

  By definition of semi-standard tableaux (which requires each integer
  to appear exactly once within the tableau $\Tilde{\Theta}$), it is
  clear that
\begin{equation}
  \label{eq:HV-Intersection}
\mathbf{h}_{\Tilde{\Theta}} \cap \mathbf{v}_{\Tilde{\Theta}} = \lbrace \mathrm{id}
\rbrace
\ ,
\end{equation}
where $\mathrm{id}$ is the identity permutation in $S_n$. 
\end{definition}

\noindent For example, if
\begin{equation}
  \label{eq:countVTheta}
  \Theta =
  \begin{ytableau}
    1 & 3 \\
    2 & 5 \\
    4
  \end{ytableau}
  \ ,
\end{equation}
then 
\begin{equation}
  \mathbf{h}_{\Theta}=\lbrace\mathrm{id},(13),(25),(13)(25)\rbrace
\end{equation}
and 
\begin{equation}
  \mathbf{v}_{\Theta}=\lbrace
\mathrm{id},(12),(14),(24),(124),(142),(35),(12)(35),(14)(35),(24)(35),(124)(35),(142)(35)
\rbrace
\ .
\end{equation}

With these definitions we can restate Lemma IV.5 of~\cite{Tung:1985na}:

\begin{lemma}[Tung's Lemma IV.5]\label{TungLemmaIV.5}
  Let $\Theta\in\mathcal{Y}_n$ be a Young tableau and let $\rho$ be a
  (linear combination of) permutation(s) in $S_n$. If $\rho$ satisfies
  \begin{equation}
    h_{\Theta} \rho v_{\Theta} = {\mathrm{sign}(v_{\Theta})} \rho
  \end{equation}
for all $h_{\Theta}\in\mathbf{h}_{\Theta}$ and for all
$v_{\Theta}\in\mathbf{v}_{\Theta}$, then $\rho$ is proportional to the
Young projection operator corresponding to $\Theta$,
\begin{equation}
  \rho = \lambda \cdot Y_{\Theta}.
\end{equation}
Furthermore, if we write $\rho$ as a sum of permutations,
\begin{equation}
  \rho = \sum_{\sigma\in S_n} a_{\sigma} \sigma,
\end{equation}
where the $a_{\sigma}$ are constants, then $\lambda$ is proportional to the
coefficient of the identity in the series expansion of $\rho$,
\begin{equation}
  \lambda = \frac{\mathcal{H}_{\Theta}}{b_{O}} a_{\mathrm{id}}
\ ,
\end{equation}
where $\mathcal{H}_{\Theta}$ denotes the hook length of $\Theta$~\cite{Sagan:2000,Fulton:1997} and
$b_O$ is the product of $(\text{length of (anti-) symmetrizer})!$ for
all symmetrizers and antisymmetrizers in $O$~\cite{Cvitanovic:2008zz}.

The last statement is not included in the original version shown
in~\cite{Tung:1985na}, but follows from the proof presented there.
\end{lemma}
It should be noted that in~\cite{Tung:1985na}, symmetrizers and
antisymmetrizers are not normalized: for example, we define
$\bm{S}_{12}:=\frac{1}{2}(\mathrm{id}+(12))$ while~\cite{Tung:1985na}
defines $\bm{S}_{12}:=\mathrm{id}+(12)$. Thus, the constant $b_{O}$
arises in our statement of the Lemma~\ref{TungLemmaIV.5}, but is not
present in~\cite{Tung:1985na}. Furthermore,~\cite{Tung:1985na}'s
statement of this Lemma compares the algebra element $\rho$ with the
\emph{irreducible symmetrizer} $e_{\Theta}$, which differs from
$Y_{\Theta}$ by the constant $\mathcal{H}_{\Theta}$, (keeping in mind
the different normalizations of symmetrizers and antisymmetrizers used
in this paper and in~\cite{Tung:1985na}). This leads to the constant
$\mathcal{H}_{\Theta}$ in our rendition of the Lemma.

Lemma~\ref{TungLemmaIV.5} immediately gives rise to the following
special case:

\begin{corollary}[Cancellation of parts of the operator]\label{thm:Cancel-Ops}
  Let $\Theta\in\mathcal{Y}_n$ be a Young tableau and $M$ an element
  of the algebra of primitive invariants
  $\in\API{\SUN,\Pow{n}}$. Then, there exists a (possibly vanishing)
  constant $\lambda$ such that
  \begin{equation}
    \label{eq:Cancel-General-O}
O := \mathbf{S}_{\Theta} \; M \; \mathbf{A}_{\Theta} =
\lambda \cdot Y_{\Theta}
\ .
  \end{equation}
If furthermore the operator $O$ is non-zero, then $\lambda\neq 0$.
\end{corollary}

Imagine that $M$ is exclusively constructed as a product of
symmetrizers and antisymmetrizers as will be the case in our
applications. Then $\Theta\in\mathcal{Y}_n$ and
$M\in\API{\SUN,\Pow{n}}$ ensures that $\mathbf{A}_{\Theta}$
is (in birdtrack parlance) the longest set of antisymmetrizers in $O$, and
$\mathbf{S}_{\Theta}$ is the longest set of symmetrizers in $O$. This
is illustrated by the following example:
\begin{equation}
\label{eq:wedged-sets-example}
  O := \;
  \scalebox{0.75}{$\underbrace{\FPic{5s354N}\FPic{5Sym123Sym45N}\FPic{5s345N}}_{\mbox{\normalsize
      $\mathbf{S}_{\Theta}$}}\underbrace{\FPic{5s23N}\FPic{5ASym12N}\FPic{5s23N}\FPic{5Sym12Sym34N}}_{\mbox{\normalsize
  $M$}}\underbrace{\FPic{5s23N}\FPic{5ASym12ASym34N}\FPic{5s23N}}_{\mbox{\normalsize
      $\mathbf{A}_{\Theta}$}}$}
\quad \text{ where } \quad
 \Theta :=
  \begin{ytableau}
    1 & 2 & 5 \\
    3 & 4
  \end{ytableau}
\ .
\end{equation} 
This observation is a key element in recognizing where
eq.~\eqref{eq:Cancel-General-O} is applicable.

\noindent\emph{Proof of Corollary~\ref{thm:Cancel-Ops}:} From the definition of horizontal and vertical permutations
(Definition~\ref{HVPermutation}) it is clear that 
\begin{IEEEeqnarray*}{rCl}
  h_{\Theta}\mathbf{S}_{\Theta}=\mathbf{S}_{\Theta} & \qquad &
  \text{for all } h_{\Theta}\in\mathbf{h}_{\Theta}
\\
 \mathbf{A}_{\Theta}v_{\Theta}={\mathrm{sign}(v_{\Theta})}\mathbf{A}_{\Theta}
 & \qquad & \text{for all } v_{\Theta}\in\mathbf v_{\Theta}
\ ,
\end{IEEEeqnarray*}
where $\mathrm{sign}(\rho)$ denotes the signature of the permutation
$\rho$.\footnote{$\mathrm{sign}(\rho)$ is $\pm 1$ depending on whether
  $\rho$ decomposes into an even or odd number of transpositions. Tung
  in~\cite{Tung:1985na} means the same when he writes
  $(-1)^{\mathrm{sign}(\rho)}$.} Since
$O:=\mathbf{S}_{\Theta}\;M\;\mathbf{A}_{\Theta}$
(eq.~\eqref{eq:Cancel-General-O}), it immediately follows that, for
all $h_{\Theta}\in\mathbf{h}_{\Theta}$ and all
$v_{\Theta}\in\mathbf{v}_{\Theta}$
\begin{IEEEeqnarray*}{rClCrCl}
  h_{\Theta}O & = & 
\raisebox{-0.3\height}{
\begin{tikzpicture}[baseline=(current bounding box.west),
  every node/.style={inner sep=1pt,outer sep=-1pt}    ]    \matrix(ID)[    matrix of math nodes,    ampersand replacement=\&,    row sep =0mm,    column sep =0mm    ]    {v_{\Theta}
      \& \mathbf{S}_{\Theta}
      \& M
      \& \mathbf{A}_{\Theta}
      \\
};
\draw[decorate,decoration={brace,amplitude=4pt},thick] (ID-1-2.south east) --
(ID-1-1.south west) node[pos=.5,anchor=north,yshift=-2mm]
{\scriptsize$\mathbf{S}_{\Theta}$};
\end{tikzpicture}
}
& \hspace{2cm} &
O v_{\Theta} & = &
\raisebox{-0.3\height}{
\begin{tikzpicture}[baseline=(current bounding box.west),
  every node/.style={inner sep=1pt,outer sep=-1pt}        ]
    \matrix(ID)[
    matrix of math nodes,
    ampersand replacement=\&,
    row sep =0mm,
    column sep =0mm
    ]
    {\mathbf{S}_{\Theta}
      \& M
      \& \mathbf{A}_{\Theta}
      \& v_{\Theta}
      \\
};
\draw[decorate,decoration={brace,amplitude=4pt},thick] (ID-1-4.south east) --
(ID-1-3.south west) node[pos=.5,anchor=north,yshift=-2mm]
{\scriptsize${\mathrm{sign}(v_{\Theta})}\mathbf{A}_{\Theta}$};
\end{tikzpicture}
}
\\
& = & 
    \mathbf{S}_{\Theta} \; 
M
\; \mathbf{A}_{\Theta} 
&& 
& = & 
{\mathrm{sign}(v_{\Theta})} \; 
    \mathbf{S}_{\Theta} \; 
M
\; \mathbf{A}_{\Theta} 
\\
& = & 
    O 
&&
& = & 
   {\mathrm{sign}(v_{\Theta})} O 
\ .
\end{IEEEeqnarray*}
More compactly, these conditions become
\begin{equation}
  \label{eq:CancelWedgedParentSetsProof5}
  h_{\Theta} O v_{\Theta} = {\mathrm{sign}(v_{\Theta})} O
  \hspace{1cm}
  \text{
  for all $h_{\Theta}\in\mathbf{h}_{\Theta}$ and
$v_{\Theta}\in\mathbf{v}_{\Theta}$.}
\end{equation}
However, according to Lemma~\ref{TungLemmaIV.5}~\cite[Lemma
IV.5]{Tung:1985na}, relation~\eqref{eq:CancelWedgedParentSetsProof5}
holds \emph{if and only if} $O$ is proportional to the Young
projection operator $Y_{\Theta}$; that is, there exists a constant
$\lambda$ such that
\begin{equation}
\label{eq:O-prop-Y}
  O = \lambda \cdot Y_{\Theta}
\ .
\end{equation}
From this, it follows immediately that
$\lambda\neq 0$ if and only if $O\neq 0$, thus establishing our
claim. \qed

One of the main cases of interest is a situation where the structure
of $O$ (and thus $M$) is such that we know from the outset
that it is nonzero. One such condition is that none of the
antisymmetrizers contained in $O$ may exceed the length $N$ -- if this
occurs we refer to it as a \emph{dimensional zero}. We will re-visit
this scenario at the end of this section.

Two further conditions ensuring $O\neq 0$ are presented below,
conditions~\ref{thm:Cancel-nonzero-O1} and~\ref{thm:Cancel-nonzero-O2}
(condition~\ref{thm:Cancel-nonzero-O3} is a combination of
conditions~\ref{thm:Cancel-nonzero-O1}
and~\ref{thm:Cancel-nonzero-O2}). We do not claim that the conditions
given in this section represent an exhaustive list of cases yielding
$O\neq 0$, but rather that these cases occur most commonly in
practical
examples,~\cite{Alcock-Zeilinger:2016sxc,Alcock-Zeilinger:2016cva}.

\begin{condition}[inclusion of (anti-) symmetrizers]\label{thm:Cancel-nonzero-O1}
  Let $O$ be of the form~\eqref{eq:Cancel-General-O}, $O =
  \mathbf{S}_{\Theta}\;M\;\mathbf{A}_{\Theta}$, and
  $M$ be given by
\begin{equation}
  \label{eq:CancelWedgedParent1}
  M = \; \mathbf{A}_{\Phi_1} \;
  \mathbf{S}_{\Phi_2} \; \mathbf{A}_{\Phi_3} \;
  \mathbf{S}_{\Phi_4} \; \cdots \; \mathbf{A}_{\Phi_{k-1}} \;
  \mathbf{S}_{\Phi_k}
\ ,
\end{equation}
such that $\mathbf{A}_{\Phi_i}\supset\mathbf{A}_{\Theta}$ for every
$i\in\lbrace 1,3,\ldots k-1\rbrace$ and
$\mathbf{S}_{\Phi_j}\supset\mathbf{S}_{\Theta}$ for every $j\in\lbrace
2,4,\ldots k\rbrace$. Then $O$ is a non-zero element of $\API{\SUN,\Pow{n}}\subset\Lin{\Pow{n}}$.
\end{condition}

\emph{Proof:} The operator $O=\mathbf{S}_{\Theta}\;M\;\mathbf{A}_{\Theta}$
with $M$ given in~\eqref{eq:CancelWedgedParent1} is defined
to be a product of alternating symmetrizers and antisymmetrizers. In
particular, the outermost sets of symmetrizers and antisymmetrizers,
$\mathbf{S}_{\Theta}$ and $\mathbf{A}_{\Theta}$ respectively,
correspond to a Young tableau $\Theta$. By the definition of Young
tableaux, this implies that each symmetrizer in $\mathbf{S}_{\Theta}$
has \emph{at most} one common leg with each antisymmetrizer in
$\mathbf{A}_{\Theta}$ (this is the underlying reason why
$\Bar Y_\Theta =\mathbf{S}_{\Theta}\mathbf{A}_{\Theta} \neq 0$).
Furthermore, since $\mathbf{S}_{\Phi_j}\supset\mathbf{S}_{\Theta}$ for
every $j\in\lbrace 2,4,\ldots k\rbrace$ and
$\mathbf{A}_{\Phi_i}\supset\mathbf{A}_{\Theta}$ for every
$i\in\lbrace 1,3,\ldots k-1\rbrace$, the same applies for every other
(not necessarily neighbouring) pair $\mathbf{S}_{\Xi_i}$ and
$\mathbf{A}_{\Xi_j}$ occurring in $O$. This guarantees that the
operator $O$ as defined in~\eqref{eq:CancelWedgedParent1} is
non-zero. \qed

As an example of condition~\ref{thm:Cancel-nonzero-O1}
consider the operator
\begin{equation}
  O = \;
  \scalebox{0.75}{$
\underbrace{
\FPic{5s354N}\FPic{5Sym123Sym45N}\FPic{5s345N}
}_{\mbox{\normalsize $\mathbf{S}_{\Theta}$}}
\underbrace{
\FPic{5s23N}\FPic{5ASym12N}\FPic{5s23N}
}_{\mbox{\normalsize $\mathbf{A}_{\Phi_1}$}}
\underbrace{\FPic{5Sym12Sym34R}
}_{\mbox{\normalsize $\mathbf{S}_{\Phi_2}$}}
\underbrace{\FPic{5s23N}\FPic{5ASym12ASym34N}\FPic{5s23N}
}_{\mbox{\normalsize $\mathbf{A}_{\Theta}$}}
$}
\ .
\end{equation}
In $O$, the sets $\mathbf{S}_{\Theta}$ and $\mathbf{A}_{\Theta}$ correspond to
the Young tableau
\begin{equation}
  \label{eq:ExWedgedAnc3}
  \Theta :=
  \begin{ytableau}
    1 & 2 & 5 \\
    3 & 4
  \end{ytableau}
\ .
\end{equation}
The inclusion conditions are
$\mathbf{A}_{\Phi_1}=\lbrace\bm{A}_{13}\rbrace\supset\lbrace\bm{A}_{13},\bm{A}_{24}\rbrace=\mathbf{A}_{\Theta}$
and
$\mathbf{S}_{\Phi_2}=\lbrace\bm{S}_{12},\bm{S}_{34}\rbrace\supset\lbrace\bm{S}_{125},\bm{S}_{34}\rbrace=\mathbf{S}_{\Theta}$.\footnote{In
  this particular case, one can even notice that the set
  $\mathbf{A}_{\Phi_2}$ corresponds to the ancestor tableau
  $\Theta_{(2)}$ and the set $\mathbf{S}_{\Phi_3}$ corresponds to the
  ancestor tableau $\Theta_{(1)}$ of $\Theta$. Hence, $Q$ can be
  written as
  $Q = \;
  \mathbf{S}_{\Theta}\mathbf{A}_{\Theta_{(2)}}\mathbf{S}_{\Theta_{(1)}}\mathbf{A}_{\Theta}$.}
Then, according to Corollary~\ref{thm:Cancel-Ops}, we may cancel the
wedged sets $\mathbf{A}_{\Phi_1}$ and $\mathbf{S}_{\Phi_2}$ at the
cost of a non-zero constant $\kappa$,
\begin{equation}
  Q = \kappa \cdot
  \scalebox{0.75}{$\underbrace{\FPic{5s345N}\FPic{5Sym123Sym45N}\FPic{5s354N}}_{\mbox{\normalsize
      $\mathbf{S}_{\Theta}$}}\underbrace{\FPic{5s23N}\FPic{5ASym12ASym34N}\FPic{5s23N}}_{\mbox{\normalsize
    $\mathbf{A}_{\Theta}$}}$}
 \; = \kappa \cdot \Bar Y_{\Theta}
\ .
\end{equation}
The simplication is noteable and nontrivial. It is useful in all
situations where the end result is simple enough and we have an
external criterion to constrain the product of any of the unknown
proportionality factors $\kappa$ acquired in the possible repeated
application of Corollary~\ref{thm:Cancel-Ops}.

A second way of constructing non-zero operators is by relating
symmetrizers and antisymmetrizers of different Young tableaux with a
permutation. To this end, we require the following definition.

\begin{definition}[tableau permutation]\label{thm:TableauPermutation}
Consider two Young tableaux $\Theta,\Phi\in\mathcal{Y}_n$ with the
same shape. Then, $\Phi$ can be obtained from $\Theta$ by permuting
the numbers of $\Theta$; clearly, the permutation needed to obtain
$\Phi$ from $\Theta$ is unique. Denote this permutation by
$\rho_{\Theta\Phi}$,
\begin{equation}
  \Theta = \rho_{\Theta\Phi} (\Phi)
\qquad \Longleftrightarrow \qquad 
  \Phi = \rho_{\Theta\Phi}^{-1} (\Theta) 
    = \rho_{\Phi\Theta} (\Theta)
\ .
\end{equation}
To construct $\rho_{\Theta\Phi}$ explicitly, write the Young
tableau $\Theta$ and $\Phi$ next to each other such that $\Theta$
is to the left of $\Phi$ and then connect the boxes in the
corresponding position of the two diagrams, such as
\begin{equation}\label{eq:SchematicTransitionPermutation}
\Theta \rightarrow \;
\diagram[scale=.6]{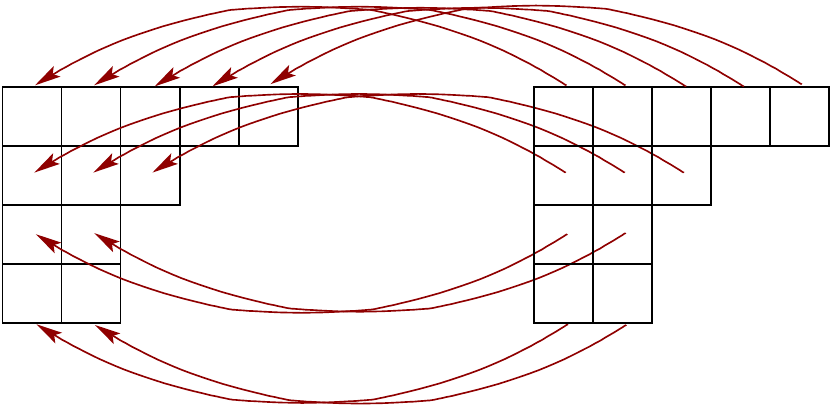} \leftarrow
\Phi.
\end{equation}
Write two columns of numbers from $1$ to $n$ next to each
other in descending order; the left column represents the entries of
$\Theta$ and the right column represents the entries of
$\Phi$. Connect the entries in the left and the right
column in correspondence to
\eqref{eq:SchematicTransitionPermutation}. The resulting tangle of
lines is the birdtrack corresponding to $\rho_{\Theta\Phi}$ and
thus determines the permutation.
\end{definition}

As an example, the permutation $\rho_{\Theta\Phi}$ between the tableaux
\begin{equation}
\Theta = 
\begin{ytableau}
  1 & 2 \\
  3
\end{ytableau}
\qquad \text{and} \qquad 
\Phi =
\begin{ytableau}
  1 & 3 \\
  2
\end{ytableau}
\end{equation}
is given by
\begin{equation}
\Theta \rightarrow
\diagram[scale=.8]{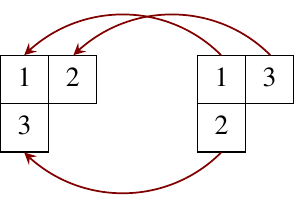}
\leftarrow \Phi
\qquad \Longrightarrow \qquad 
\rho_{\Theta\Phi} = \; \FPic{3s23SN}
\ .
\end{equation}

Let $\Theta$ and $\Phi$ be two Young tableaux of the same shape and
construct the permutation $\rho_{\Theta\Phi}$. Furthermore, consider a
general operator $K_{\Theta}$ comprised of sets of (anti-)
symmetrizers which can be absorbed into $\mathbf{S}_{\Theta}$ and
$\mathbf{A}_{\Theta}$ respectively, and let $H_{\Phi}$ be an operator
comprised of sets of (anti-) symmetrizers which can be absorbed into
$\mathbf{S}_{\Phi}$ and $\mathbf{A}_{\Phi}$ respectively.  Except for
isolated examples, the product $K_{\Theta}\cdot H_{\Phi}$
vanishes.\footnote{This is true since the product of (most!) Young
  projection operators corresponding to different Young tableaux of
  the same shape in $\mathcal{Y}_n$
  vanishes~\cite{Littlewood:1950,Tung:1985na}.} However, it turns out
that
\begin{equation}
\label{eq:ThetaThetap-nonzero-product}
  H_{\Phi}\cdot \underbrace{\rho_{\Theta\Phi}^{-1}}_{\rho_{\Phi\Theta}}
  K_{\Theta}\rho_{\Theta\Phi} \neq 0
\qquad 
\text{for all $\Theta,\Phi\in\mathcal{Y}_n$ for all $n$}
\ .
\end{equation}
To better understand this, we accompany the general argument with an example: Consider the Young tableaux
 \begin{equation}
\label{eq:ThetaThetap-Tableaux}
   \Theta = \begin{ytableau}
     1 & 3 & 5 \\
     2 & 4 \\
     6
   \end{ytableau} \qquad \text{and} \qquad
   \Phi = \begin{ytableau}
     1 & 2 & 6 \\
     3 & 5 \\
     4
   \end{ytableau}
\ .
 \end{equation}
 The permutation $\rho_{\Theta\Phi}$ as defined
 in Definition~\ref{thm:TableauPermutation} is given by
 \begin{equation}
   \label{eq:ThetaPhi-Rho}
   \rho_{\Theta\Phi} = \scalebox{0.75}{\FPic{6s23s465}}
\ .
 \end{equation}
 For a general Young tableau $\Psi\in\mathcal{Y}_n$, we denote the
 irregular tableau that is obtained from $\Psi$ by deleting the boxes
 with entries $a_1$ up to $a_m$ ($m\leq n$) by
 $\Psi\setminus\lbrace a_1,\ldots a_m\rbrace$. Even though
 $\Psi\setminus\lbrace a_1,\ldots a_m\rbrace$ is not a Young tableau
 in general, it remains semi-standard. Thus, the (anti-) symmetrizers
 in the sets
 $\mathbf{S}_{\Psi\setminus\lbrace a_1,\ldots a_m\rbrace}$ and
 $\mathbf{A}_{\Psi\setminus\lbrace a_1,\ldots a_m\rbrace}$ are
 disjoint and the sets themselves individually remain Hermitian projection
 operators. These 
 sets can further be
 absorbed into $\mathbf{S}_{\Psi}$ and $\mathbf{A}_{\Psi}$ respectively since
 $\mathbf{S}_{\Psi\setminus\lbrace a_1,\ldots a_m\rbrace}$ is merely
 the set of symmetrizers $\mathbf{S}_{\Psi}$ with the legs $a_1$ up to $a_m$
 deleted, and similarly for
 $\mathbf{A}_{\Psi\setminus\lbrace a_1,\ldots a_m\rbrace}$. Thus, they
 satisfy the absorbtion relations
 \begin{equation}
\label{eq:Absorb-Fragments}
   \mathbf{S}_{\Psi\setminus\lbrace a_1,\ldots a_m\rbrace}\mathbf{S}_{\Psi} 
= \mathbf{S}_{\Psi} 
= \mathbf{S}_{\Psi}\mathbf{S}_{\Psi\setminus\lbrace a_1,\ldots a_m\rbrace}
\hspace{1cm} \text{and} \hspace{1cm}
   \mathbf{A}_{\Psi\setminus\lbrace a_1,\ldots a_m\rbrace}\mathbf{A}_{\Psi} 
= \mathbf{A}_{\Psi} 
= \mathbf{A}_{\Psi}\mathbf{A}_{\Psi\setminus\lbrace a_1,\ldots
  a_m\rbrace}
\ ,
 \end{equation}
this is easiest seen via the birdtracks corresponding to the
semi-standard irregular tableau $\Psi\setminus\lbrace a_1,\ldots a_m\rbrace$.

A quick look at our example elucidates how
equation~\eqref{eq:Absorb-Fragments} comes about in general:
In~\eqref{eq:ThetaThetap-Tableaux}, we may remove boxes from $\Theta$
at will -- consider for example
\begin{IEEEeqnarray}{rCCCCCl}
& \Theta &
& \Theta &
& \Theta & \nonumber \\
  & \overbrace{\begin{ytableau}
     1 & *(cyan!40) 3 & 5 \\
     2 & 4 \\
     *(cyan!40) 6
   \end{ytableau}} &
& \overbrace{\begin{ytableau}
     1 & 3 & *(cyan!40) 5 \\
     2 & *(cyan!40) 4 \\
     *(cyan!40) 6
   \end{ytableau}} &
& \overbrace{\begin{ytableau}
     1 & 3 & 5 \\
     *(cyan!40) 2 & 4 \\
     6
   \end{ytableau}} & \nonumber \\
& \FPic{StraightDownArrow} & \hspace{2cm} & \FPic{StraightDownArrow} &
 \hspace{2cm} & \FPic{StraightDownArrow} & \nonumber \\
  & \underbrace{\begin{ytableau}
     1 & \none & 5 \\
     2 & 4 
   \end{ytableau}} &
& \underbrace{\begin{ytableau}
     1 & 3 \\
     2
   \end{ytableau}} &
& \underbrace{\begin{ytableau}
     1 & 3 & 5 \\
    \none & 4 \\
     6
   \end{ytableau}} & \ . \label{eq:DeleteTheta} \\
& \Theta\setminus\lbrace 3,6\rbrace &
& \Theta\setminus\lbrace 4,5,6\rbrace &
& \Theta\setminus\lbrace 2\rbrace & \nonumber 
\end{IEEEeqnarray}
It is clear from this list that only some of the resulting tableaux
will be Young tableaux, most will not. Using the tableaux~\eqref{eq:DeleteTheta}, we 
construct an operator $K_{\Theta}$ consisting of (anti-) symmetrizers
which can be absorbed into $\mathbf{S}_{\Theta}$ and
$\mathbf{A}_{\Theta}$,
\begin{align}
  K_{\Theta} & := \mathbf{S}_{\Theta\setminus\lbrace
    3,6\rbrace}\mathbf{A}_{\Theta\setminus\lbrace
    2\rbrace}\mathbf{S}_{\Theta}\mathbf{A}_{\Theta}\mathbf{S}_{\Theta\setminus\lbrace
    2\rbrace}\mathbf{A}_{\Theta\setminus\lbrace 3,6\rbrace}\mathbf{S}_{\Theta\setminus\lbrace
    4,5,6\rbrace} \nonumber \\
& = \scalebox{0.75}{\FPic{6s253SN}\FPic{6Sym12Sym34N}\FPic{6s26s35SN}\FPic{6ASym12ASym34N}\FPic{6s236s45SN}\FPic{6Sym123Sym45N}\FPic{6s24s36SN}\FPic{6ASym123ASym45N}\FPic{6s24s36SN}\FPic{6Sym123N}\FPic{6s2453SN}\FPic{6ASym12N}\FPic{6s23SN}\FPic{6Sym12N}\FPic{6s23SN}}
\ .
\end{align}
Conjugating the operator $K_{\Theta}$ by the permutation
$\rho_{\Theta\Phi}$ yields
\begin{equation}
  \label{eq:KTheta-Rho-Transform}
  \underbrace{\scalebox{0.75}{\FPic{6s23s456}}}_{\rho_{\Phi\Theta}}
\underbrace{\scalebox{0.75}{\FPic{6s253SN}\FPic{6Sym12Sym34N}\FPic{6s26s35SN}\FPic{6ASym12ASym34N}\FPic{6s236s45SN}\FPic{6Sym123Sym45N}\FPic{6s24s36SN}\FPic{6ASym123ASym45N}\FPic{6s24s36SN}\FPic{6Sym123N}\FPic{6s2453SN}\FPic{6ASym12N}\FPic{6s23SN}\FPic{6Sym12N}\FPic{6s23SN}
}}_{K_{\Theta}}
\underbrace{\scalebox{0.75}{\FPic{6s23s465}}}_{\rho_{\Theta\Phi}}
\ ;
\end{equation}
Each of the sets of (anti-) symmetrizers in
\eqref{eq:KTheta-Rho-Transform} corresponds to one of the tableaux
\begin{IEEEeqnarray}{rCCCCCl}
& \Phi &
& \Phi &
& \Phi & \nonumber \\
  & \overbrace{\begin{ytableau}
     1 & *(cyan!40) 2 & 6 \\
     3 & 5 \\
     *(cyan!40) 4
   \end{ytableau}} &
& \overbrace{\begin{ytableau}
     1 & 2 & *(cyan!40) 6 \\
     3 & *(cyan!40) 5 \\
     *(cyan!40) 4
   \end{ytableau}} &
& \overbrace{\begin{ytableau}
     1 & 2 & 6 \\
     *(cyan!40) 3 & 5 \\
     4
   \end{ytableau}} & \nonumber \\
& \FPic{StraightDownArrow} & \hspace{2cm} & \FPic{StraightDownArrow} &
 \hspace{2cm} & \FPic{StraightDownArrow} &\nonumber \\
  & \underbrace{\begin{ytableau}
     1 & \none & 6 \\
     3 & 5 
   \end{ytableau}} &
& \underbrace{\begin{ytableau}
     1 & 2 \\
     3
   \end{ytableau}} &
& \underbrace{\begin{ytableau}
     1 & 2 & 6 \\
    \none & 5 \\
     4
   \end{ytableau}} \ .& \label{eq:DeleteThetap} \\
& \Phi\setminus\lbrace 2,4\rbrace &
& \Phi\setminus\lbrace 4,5,6\rbrace &
& \Phi\setminus\lbrace 3\rbrace & \nonumber 
\end{IEEEeqnarray}
The tableaux in~\eqref{eq:DeleteThetap} are obtained by 
superimposing the tableaux in~\eqref{eq:DeleteTheta} on $\Phi$ in a cookie
cutter fashion. By construction, all the
$\mathbf{S}_{\Phi\setminus\lbrace b_1,\ldots b_m\rbrace}$
(resp. $\mathbf{A}_{\Phi\setminus\lbrace b_1,\ldots b_m\rbrace}$) 
can be absorbed into 
$\mathbf{S}_{\Phi}$ (resp. $\mathbf{A}_{\Phi}$), as claimed in eq.~\eqref{eq:Absorb-Fragments}.

The pattern is completely general and in no way restricted to the
particular example used to demonstrate it. Let us summarize:

\begin{condition}[relating (anti-) symmetrizers across tableaux]\label{thm:Cancel-nonzero-O2}
  Let $O$ be of the form
  $O=\mathbf{S}_{\Theta}\;M\;\mathbf{A}_{\Theta}$,
  eq.~\eqref{eq:Cancel-General-O}. Let
  $\Theta,\Phi\in\mathcal{Y}_n$ be two Young tableaux with the same
  shape and construct the permutation $\rho_{\Theta\Phi}$ between
  the two tableaux according to Definition~\ref{thm:TableauPermutation}. Furthermore, let $\mathcal{D}_{\Theta}$ be a product of
  symmetrizers and antisymmetrizers, each of which can be absorbed into
  $\mathbf{S}_{\Theta}$ and $\mathbf{A}_{\Theta}$ respectively.
 If $M$ is of the form
  \begin{equation}
    M = \rho_{\Phi\Theta}\mathcal{D}_{\Theta} \; \rho_{\Theta\Phi}
    \ ,
  \end{equation}
then the operator $O$ is non-zero.
\end{condition}

It immediately follows that a combination of
conditions~\ref{thm:Cancel-nonzero-O1} and~\ref{thm:Cancel-nonzero-O2}
also renders the operator $O$ non-zero:

\begin{condition}[combining conditions~\ref{thm:Cancel-nonzero-O1}
and~\ref{thm:Cancel-nonzero-O2}]\label{thm:Cancel-nonzero-O3}
  Let $O$ be an operator of the form
  $O=\mathbf{S}_{\Theta}\;M\;\mathbf{A}_{\Theta}$ and let $M$ be given
  by
  \begin{equation}
    M = M^{(1)} \; M^{(2)} \cdots M^{(l)},
  \end{equation}
such that for each $M^{(i)}$ either condition~\ref{thm:Cancel-nonzero-O1}
or~\ref{thm:Cancel-nonzero-O2} holds; this implies that each (anti-)
symmetrizer in $M$ can be absorbed into $\mathbf{S}_{\Theta}$ or
$\mathbf{A}_{\Theta}$ respectively. Then $O$ is nonzero.
\end{condition}

\paragraph{Dimensional zeroes:} Let us conclude this section with a
short discussion on how the operator $O$ becomes dimensionally
zero. Since in either of the three conditions presented in this
section all sets of antisymmetrizers in $M$ can be absorbed into
$\mathbf{A}_{\Theta}$, 
\begin{equation}
\label{eq:Dimensional-Zero-Absorb}
  \mathbf{A}_{j} \mathbf{A}_{\Theta} = \mathbf{A}_{\Theta} = 
  \mathbf{A}_{\Theta} \mathbf{A}_{j}
\ ,
\end{equation}
for every $\mathbf{A}_{j}$ in $M$, it
follows immediately that the antisymmetrizer in $O$ that contains the
most legs (i.e. the ``longest'' antisymmetrizer in $O$) must be part
of the set $\mathbf{A}_{\Theta}$, as otherwise
eq.~\eqref{eq:Dimensional-Zero-Absorb} could not hold. Thus, $O$ is
not dimensionally zero if $\mathbf{A}_{\Theta}$ is not dimensionally
zero. Furthermore, since
$Y_{\Theta}\propto\mathbf{S}_{\Theta}\mathbf{A}_{\Theta}$, it suffices
to require that $N$ is large enough for the Young projection operator
$Y_{\Theta}$ to be non-zero to ensure that the operator $O$ in any of
the
conditions~\ref{thm:Cancel-nonzero-O1}--\ref{thm:Cancel-nonzero-O3} is
not dimensionally zero. Thus, in cancelling parts of the operator $O$ (to
give it the structural form of $Y_{\Theta}$), one does not remove any
indication of it being dimensionally zero: dimensional zeroes of $O$
occur exactly when $Y_{\Theta}$ is zero.

The cancellation rules given in this section are of enormous practical
use as they allow us to shorten birdtrack operators, often
significantly so.  In particular, we use
Corollary~\ref{thm:Cancel-Ops} in the construction of compact
Hermitian Young projection
operators~\cite{Alcock-Zeilinger:2016sxc} and the
construction of transition
operators~\cite{Alcock-Zeilinger:2016cva}.

\section{Propagation rules}\label{sec:PropagateRules}

In this section, we will present propagation rules that allow us to
propagate certain symmetrizers through sets of antisymmetrizers, and
vice versa. These rules are particularly useful to make the Hermiticity of
certain birdtrack operators visually explicit. We demonstrate their
effectiveness in~\cite{Alcock-Zeilinger:2016sxc} and
with a specific example in our conclusions, see Fig.~\ref{fig:MOLDAdvantage}.

The structure of our proof of these propagation rules has been
strongly inspired by an example presented in the Appendix of Keppeler
and Sj{\"o}dahl's (KS) paper on Hermitian Young projection
operators~\cite{Keppeler:2013yla}. In this example, KS clearly
realized that symmetrizers sometimes can be propagated through sets of
antisymmetrizers, and vice versa, by ``swapping'' appropriate sets of
antisymmetrizers around; however, the general conditions under which
this is possible were not identified by KS, and a proof is also not
present in~\cite{Keppeler:2013yla}.

\begin{theorem}[propagation of (anti-) symmetrizers]
  \label{thm:PropagateSyms}
  Let $\Theta$ be a Young tableau and $O$ be a birdtrack operator of
  the form
  \begin{equation}
    \label{eq:PropagateSyms1}
    O = \mathbf{S}_{\Theta} \; \mathbf{A}_{\Theta} \; 
    \mathbf{S}_{\Theta\setminus\mathcal R},
  \end{equation}
  in which the symmtrizer set $\mathbf{S}_{\Theta\setminus\mathcal R}$
  arises from $\mathbf{S}_{\Theta}$ by removing precisely one
  symmetrizer $\bm{S}_{\mathcal R}$. By definition $\bm{S}_{\mathcal
    R}$ corresponds to a row $\mathcal R$ in $\Theta$ such that
  \begin{equation}
    \label{eq:STheta-SR}
\mathbf{S}_{\Theta} = \mathbf{S}_{\Theta\setminus\mathcal R}
  \bm{S}_{\mathcal R} = \bm{S}_{\mathcal R}
  \mathbf{S}_{\Theta\setminus\mathcal R}
\ .
  \end{equation}

  If the column-amputated tableau of $\Theta$ according to the row
  $\mathcal{R}$, $\cancel{\Theta}_c\left[\mathcal{R}\right]$, is
  \textbf{rectangular}, then the symmetrizer $\bm{S}_{\mathcal R}$ may be
  propagated through the set $\mathbf{A}_{\Theta}$ from the left to
  the right, yielding
  \begin{equation}
    O=\mathbf{S}_{\Theta} \; \mathbf{A}_{\Theta} \;
    \mathbf{S}_{\Theta\setminus\mathcal R}= \mathbf{S}_{\Theta\setminus\mathcal R} \; \mathbf{A}_{\Theta} \;
    \mathbf{S}_{\Theta}
    \ ,
  \end{equation}
  which implies that $O$ is Hermitian.\footnote{Recall the Hermiticity of
    (sets of) (anti-) symmetrizers, eq.~\eqref{eq:ASymsHC}.} We may think of this procedure
  as moving the missing symmetrizer $\bm{S}_{\mathcal R}$ through the
  intervening antisymmetrizer set
  $\mathbf{A}_{\Theta}$. Eq.~\eqref{eq:STheta-SR} immediately allows
  us to augment this statement to
  \begin{equation}
    \label{eq:PropagateSyms2}
    \mathbf{S}_{\Theta} \; \mathbf{A}_{\Theta} \;
    \mathbf{S}_{\Theta\setminus\mathcal R}= \mathbf{S}_{\Theta\setminus\mathcal R} \; \mathbf{A}_{\Theta} \;
    \mathbf{S}_{\Theta}
    =
    \mathbf{S}_{\Theta} \; \mathbf{A}_{\Theta} \;\mathbf{S}_{\Theta}
    \ .
  \end{equation}

  If the roles of symmetrizers and antisymmetrizers are exchanged, we
  need to verify that the row-amputated tableau
  $\cancel{\Theta}_r\left[\mathcal{C}\right]$ with respect to a column
  $\mathcal C$ is rectangular to guarantee that
  \begin{equation}
    \label{eq:PropagateSyms3}
    \mathbf{A}_{\Theta} \; \mathbf{S}_{\Theta} \;
     \mathbf{A}_{\Theta\setminus\mathcal C}= \mathbf{A}_{\Theta\setminus\mathcal C} \; \mathbf{S}_{\Theta} \;
    \mathbf{A}_{\Theta} 
    =
    \mathbf{A}_{\Theta} \; \mathbf{S}_{\Theta} \;\mathbf{A}_{\Theta}
    \ .
  \end{equation}
  This amounts to moving the missing antisymmetrizer $\bm{A}_{\mathcal
    C}$ through the intervening symmetrizer set $\mathbf{S}_{\Theta}$.
\end{theorem}

\noindent To clarify the statement of the 
Propagation-Theorem~\ref{thm:PropagateSyms}, consider for example the operator $O$
\begin{equation}
  \label{eq:SimplificationEx1}
  O := \;
  \scalebox{0.7}{$\underbrace{\FPic{7Sym123Sym45Sym67R}}_{\mbox{\large$\mathbf{S}_{\Theta}$}}\underbrace{\FPic{7s24s367}\FPic{7ASym123ASym456N}\FPic{7s24s376}}_{\mbox{\large$\mathbf{A}_{\Theta}
$}}\underbrace{\FPic{7Sym123Sym45R}}_{\mbox{\large$\mathbf{S}_{\Theta\setminus\mathcal{R}}$}}$}
\ ,
\end{equation}
where the Young tableau $\Theta$ is
\begin{equation}
  \Theta = 
  \begin{ytableau}
    1 & 2 & 3 \\
    4 & 5 \\
    6 & 7
  \end{ytableau}
\ .
\end{equation}
The operator~\eqref{eq:SimplificationEx1} meets the conditions laid out in Theorem
\ref{thm:PropagateSyms}: The sets $\mathbf{S}_{\Theta}$ and
$\mathbf{S}_{\Theta\setminus\mathcal{R}}$ differ only by one symmetrizer, namely
$\bm{S}_{\mathcal{R}}=\bm{S}_{67}$, which corresponds to the row $(6,7)$ of the tableau $\Theta$. Indeed, we find that the
amputated tableau $\cancel{\Theta}_c\left[(6,7)\right]$ is rectangular,
\begin{equation}
  \cancel{\Theta}_c\left[(6,7)\right] = 
  \begin{ytableau}
    1 & 2 \\
    4 & 5 \\
    *(blue!25) 6 & *(blue!25) 7
  \end{ytableau}
\ ,
\end{equation}
where we have highlighted the row corresponding to the symmetrizer
$\bm{S}_{67}$ in blue. We therefore may commute the symmetrizer
$\bm{S}_{67}$ from the left of $O$ to the right in accordance with the
Propagation-Theorem~\ref{thm:PropagateSyms},
\begin{equation}
  O := \;
  \scalebox{0.7}{$\FPic{7Sym123Sym45Sym67N}\FPic{7s24s367}\FPic{7ASym123ASym456N}\FPic{7s24s376}\FPic{7Sym123Sym45N}$}
    \; = \;
    \scalebox{0.7}{$\FPic{7Sym123Sym45N}\FPic{7s24s367}\FPic{7ASym123ASym456N}\FPic{7s24s376}\FPic{7Sym123Sym45Sym67N}$}
\ .
\end{equation}
Furthermore, if we factor the symmetrizer $\bm{S}_{67}$ out of the set
$\mathbf{S}_{\Theta}$ (i.e. if we write
$\mathbf{S}_{\Theta}=\bm{S}_{67}\mathbf{S}_{\Theta}$) before commuting it through, we obtain 
\begin{equation}
\label{eq:SimplificationEx1-Herm}
  O := \; \scalebox{0.7}{$\FPic{7Sym123Sym45Sym67N}\FPic{7s24s367}\FPic{7ASym123ASym456N}\FPic{7s24s376}\FPic{7Sym123Sym45N}$}
    \; = \;
\scalebox{0.7}{$   \FPic{7Sym67N}\FPic{7Sym123Sym45Sym67N}\FPic{7s24s367}\FPic{7ASym123ASym456N}\FPic{7s24s376}\FPic{7Sym123Sym45N}$}
    \; \xlongequal{\text{\tiny Thm.~\ref{thm:PropagateSyms}}} \;
 \scalebox{0.7}{$ \FPic{7Sym67N}\FPic{7Sym123Sym45N}\FPic{7s24s367}\FPic{7ASym123ASym456N}\FPic{7s24s376}\FPic{7Sym123Sym45Sym67N}$}
\; = \;
\scalebox{0.7}{$\FPic{7Sym123Sym45Sym67N}\FPic{7s24s367}\FPic{7ASym123ASym456N}\FPic{7s24s376}\FPic{7Sym123Sym45Sym67N}$}
\ .
\end{equation}
We thus kept $\bm{S}_{67}$ on both sides of the operator, making the
Hermiticity of $O$ explicit. 

Having understood the statement of the
Propagation-Theorem~\ref{thm:PropagateSyms}, we will foreshadow the
strategy of the proof (which is given in
section~\ref{sec:PropagateProofSection}). Consider the operator
\begin{equation}
  \label{eq:HermitizeResultOp}
P := \;
\FPic{4Sym12Sym34N}\FPic{4s23N}\FPic{4ASym12ASym34N}\FPic{4s23N}\FPic{4Sym12N}
\ ,
\end{equation}
which satisfies all conditions posed in the Propagation-Theorem
\ref{thm:PropagateSyms}. It thus immediately follows from the Theorem that 
\begin{equation}
\label{eq:HermitizeResult}
 \FPic{4Sym12Sym34N}\FPic{4s23N}\FPic{4ASym12ASym34N}\FPic{4s23N}\FPic{4Sym12N}
 \; = \; 
\FPic{4Sym12Sym34N}\FPic{4s23N}\FPic{4ASym12ASym34N}\FPic{4s23N}\FPic{4Sym12Sym34N}
 \; = \; 
\FPic{4Sym12N}\FPic{4s23N}\FPic{4ASym12ASym34N}\FPic{4s23N}\FPic{4Sym12Sym34N}
\ .
\end{equation}
We would however like to show how this comes about explicitly, thus
alluding to the strategy used in the proof of Theorem
\ref{thm:PropagateSyms}. In
particular, we will use a trick originally used by Keppeler and
Sj{\"o}dahl in the appendix of~\cite{Keppeler:2013yla}.

We begin by factoring a transposition out of each symmetrizer on the
left; this will not alter the operator $P$ in any way since
\begin{equation}
  \FPic{2Sym12SN}
\; = \; 
\FPic{2s12SN}\FPic{2Sym12N}
\; = \; 
\FPic{2Sym12N}\FPic{2s12SN}
\ .
\end{equation}
We thus have that
\begin{IEEEeqnarray}{rCl}
& 
\scalebox{0.75}{{\color{red}\circled{1}}} & \nonumber \\
P = \; 
\FPic{4Sym12Sym34N}\FPic{4s23N}\FPic{4ASym12ASym34N}\FPic{4s23N}\FPic{4Sym12N}\; = \;
\FPic{4Sym12Sym34N}{\setlength{\fboxsep}{1pt}\colorbox{red!20}{\FPic{4s12s34N}}}\FPic{4s23N}&
\FPic{4ASym12ASym34N}&
\FPic{4s23N}\FPic{4Sym12N}\ , \\
& \scalebox{0.75}{{\color{red}\circled{2}}} & \nonumber
\end{IEEEeqnarray}
where we have marked the top and bottom antisymmetrizer in $P$ as
\scalebox{0.85}{\color{red}\circled{1}} and \scalebox{0.85}{\color{red}\circled{2}} respectively. It
is important to notice that these two antisymmetrizers would be
indistinguishable if it weren't for the labelling. We may thus
exchange them (paying close attention to which line enters and exits
which antisymmetrizer), without changing the operator P,
\begin{IEEEeqnarray}{rClCrCl}
  & \scalebox{0.75}{{\color{red}\circled{1}}} & & & &
      \scalebox{0.75}{{\color{red}\circled{2}}} & \nonumber \\
\FPic{4Sym12Sym34N}{\setlength{\fboxsep}{1pt}\colorbox{red!20}{\FPic{4s12s34N}}}\FPic{4s23N}& 
\FPic{4ASym12ASym34N}&
\FPic{4s23N}\FPic{4Sym12N}\quad
& \; \xlongequal[]{\scalebox{0.6}{\color{red}\circled{1}} 
   \; \leftrightarrow \;
   \scalebox{0.6}{\color{red}\circled{2}} } \; & 
\quad
\FPic{4Sym12Sym34N}\FPic{4s23N}&
\FPic{4ASym12ASym34N}&
\FPic{4s23N}{\setlength{\fboxsep}{1pt}\colorbox{red!20}{\FPic{4s12s34N}}}\FPic{4Sym12N}\ . \label{eq:HermitizeStep3}
\\
& \scalebox{0.75}{{\color{red}\circled{2}}} & & & &
   \scalebox{0.75}{{\color{red}\circled{1}}} & \nonumber
\end{IEEEeqnarray}
We have thus effectively commuted the transpositions marked in red through the set of
antisymmetrizers from the left to the right. We may now absorb
the transposition on top into the right symmetrizer,
\begin{equation}
\FPic{4Sym12Sym34N}\FPic{4s23N}\FPic{4ASym12ASym34N}\FPic{4s23N}{\setlength{\fboxsep}{1pt}\colorbox{red!20}{\FPic{4s12s34N}}}\FPic{4Sym12N}   \; = \; 
\FPic{4Sym12Sym34N}\FPic{4s23N}\FPic{4ASym12ASym34N}\FPic{4s23N}{\setlength{\fboxsep}{1pt}\colorbox{red!20}{\FPic{4Sym12s34N}}}\ .
\end{equation}
We therefore showed that 
\begin{equation}
  \label{eq:HermitizeStep4}
P = \; 
\FPic{4Sym12Sym34N}\FPic{4s23N}\FPic{4ASym12ASym34N}\FPic{4s23N}\FPic{4Sym12N}   \; = \; 
\FPic{4Sym12Sym34N}\FPic{4s23N}\FPic{4ASym12ASym34N}\FPic{4s23N}\FPic{4Sym12s34N}\ .
\end{equation}
It now remains to add up the two different expressions of $P$ found
in~\eqref{eq:HermitizeStep4}, and multiply this sum by a factor $\nicefrac{1}{2}$,
\begin{equation}
  \label{eq:HermitizeStep5}
  \underbrace{\frac{1}{2}\left( \FPic{4Sym12Sym34N}\FPic{4s23N}\FPic{4ASym12ASym34N}\FPic{4s23N}\FPic{4Sym12N}
\; + \;
\FPic{4Sym12Sym34N}\FPic{4s23N}\FPic{4ASym12ASym34N}\FPic{4s23N}\FPic{4Sym12N}
\right)}_{= \;
\scalebox{0.8}{
\FPic{4Sym12Sym34N}\FPic{4s23N}\FPic{4ASym12ASym34N}\FPic{4s23N}\FPic{4Sym12N}
}
\; = \; P }
\;  = \; \frac{1}{2} \left(
\FPic{4Sym12Sym34N}\FPic{4s23N}\FPic{4ASym12ASym34N}\FPic{4s23N}\FPic{4Sym12s34N}
\; + \;
\FPic{4Sym12Sym34N}\FPic{4s23N}\FPic{4ASym12ASym34N}\FPic{4s23N}\FPic{4Sym12N} \right).
\end{equation}
However, since
\begin{equation}
  \frac{1}{2} \left( 
\FPic{4Sym12SN}
\; + \; 
\FPic{4Sym12s34SN}
\right) 
\; = \; 
\FPic{4Sym12Sym34SN}
\ ,
\end{equation}
equation~\eqref{eq:HermitizeStep5} simply becomes
\begin{equation}
  P = \; 
\FPic{4Sym12Sym34N}\FPic{4s23N}\FPic{4ASym12ASym34N}\FPic{4s23N}\FPic{4Sym12Sym34N}\ .
\end{equation}
Performing the above process in reverse then yields
\begin{equation}
  P = \;
 \FPic{4Sym12Sym34N}\FPic{4s23N}\FPic{4ASym12ASym34N}\FPic{4s23N}\FPic{4Sym12N}
  \; = \;
\FPic{4Sym12Sym34N}\FPic{4s23N}\FPic{4ASym12ASym34N}\FPic{4s23N}\FPic{4Sym12Sym34N}
  \; = \; 
\FPic{4Sym12N}\FPic{4s23N}\FPic{4ASym12ASym34N}\FPic{4s23N}\FPic{4Sym12Sym34N}
\ ,
\end{equation}
as desired. That this strategy can be applied to the
operator~\eqref{eq:SimplificationEx1} can be seen via factoring out a
symmetrizer of length $2$ from each symmetrizer,
\begin{equation}
\label{eq:SimplificationEx1-TildeO}
\begin{tikzpicture}[baseline=(current bounding box.west),
  every node/.style={inner sep=0pt,outer sep=0pt}        ]
    \matrix(ID)[
    matrix of math nodes,
    ampersand replacement=\&,
    row sep =0mm,
    column sep =0mm
    ]
    { O = \; 
      \& \scalebox{0.75}{\FPic{7Sym123Sym45Sym67N}}
      \& \scalebox{0.75}{\FPic{7s24s367}}
      \& \scalebox{0.75}{\FPic{7ASym123ASym456N}}
      \& \scalebox{0.75}{\FPic{7s24s376}}
      \& \scalebox{0.75}{\FPic{7Sym123Sym45N}}
      \& \; = \;
      \& \scalebox{0.75}{\FPic{7Sym123Sym45Sym67N}}
      \& \scalebox{0.75}{\FPic{7Sym12Sym45Sym67N}}
      \& \scalebox{0.75}{\FPic{7s24s367}}
      \& \scalebox{0.75}{\FPic{7ASym123ASym456N}}
      \& \scalebox{0.75}{\FPic{7s24s376}}
      \& \scalebox{0.75}{\FPic{7Sym12Sym45N}}
      \& \scalebox{0.75}{\FPic{7Sym123Sym45N}}
      \ .
      \\
};
\draw[draw=none,fill=blue!20,thick] ($(ID-1-9.north west) + (0pt,2pt)$) rectangle
($(ID-1-13.south east) + (0pt,-2pt)$) node[midway] {$\scalebox{0.75}{\FPic{7Sym12Sym45Sym67N}\FPic{7s24s367}\FPic{7ASym123ASym456N}\FPic{7s24s376}\FPic{7Sym12Sym45N}}$};
\draw[decorate,decoration={brace,amplitude=5pt},thick] ($(ID-1-13.south east) + (0pt,-2pt)$) --
($(ID-1-9.south west) + (0pt,-2pt)$) node[pos=.5,anchor=north,yshift=-2.5mm] {$=:\tilde{O}$};
\draw[-stealth,thick,red] (ID-1-8.north) |- ++(0,3mm)   -|
(ID-1-9.north) node[pos=.2,anchor=south,yshift=1mm] {\scriptsize factor};
\draw[stealth-,thick,red] (ID-1-13.north) |- ++(0,3mm)   -|
(ID-1-14.north) node[pos=.2,anchor=south,yshift=1mm] {\scriptsize
  factor};
\end{tikzpicture}
\end{equation}
The part marked $\tilde{O}$ in~\eqref{eq:SimplificationEx1-TildeO} can
now be dealt with exactly as in the previous example, allowing one to
commute the symmetrizer $\bm{S}_{67}$ from the left to the right. It remains to reabsorb the extra symmetrizers to obtain the desired
result~\eqref{eq:SimplificationEx1-Herm},
\begin{equation}
  \begin{tikzpicture}[baseline=(current bounding box.west),
  every node/.style={inner sep=0pt,outer sep=0pt}        ]
    \matrix(ID)[
    matrix of math nodes,
    ampersand replacement=\&,
    row sep =0mm,
    column sep =0mm
    ]
    { O = \; 
      \& \scalebox{0.75}{\FPic{7Sym123Sym45Sym67N}}
      \& \scalebox{0.75}{\FPic{7Sym12Sym45Sym67N}}
      \& \scalebox{0.75}{\FPic{7s24s367}}
      \& \scalebox{0.75}{\FPic{7ASym123ASym456N}}
      \& \scalebox{0.75}{\FPic{7s24s376}}
      \& \scalebox{0.75}{\FPic{7Sym12Sym45N}}
      \& \scalebox{0.75}{\FPic{7Sym123Sym45N}}
      \& \; = \;
      \& \scalebox{0.75}{\FPic{7Sym123Sym45Sym67N}}
      \& \scalebox{0.75}{\FPic{7Sym12Sym45Sym67N}}
      \& \scalebox{0.75}{\FPic{7s24s367}}
      \& \scalebox{0.75}{\FPic{7ASym123ASym456N}}
      \& \scalebox{0.75}{\FPic{7s24s376}}
      \& \scalebox{0.75}{\FPic{7Sym12Sym45Sym67N}}
      \& \scalebox{0.75}{\FPic{7Sym123Sym45N}}
      \& \; = \;
      \& \scalebox{0.75}{\FPic{7Sym123Sym45Sym67N}}
      \& \scalebox{0.75}{\FPic{7s24s367}}
      \& \scalebox{0.75}{\FPic{7ASym123ASym456N}}
      \& \scalebox{0.75}{\FPic{7s24s376}}
      \& \scalebox{0.75}{\FPic{7Sym123Sym45Sym67N}}
      \ .
      \\
};
\draw[draw=none,fill=blue!20,thick] ($(ID-1-3.north west) + (0pt,2pt)$) rectangle
($(ID-1-7.south east) + (0pt,-2pt)$) node[midway] {$
\scalebox{0.75}{\FPic{7Sym12Sym45Sym67N}\FPic{7s24s367}\FPic{7ASym123ASym456N}\FPic{7s24s376}\FPic{7Sym12Sym45N}}
$};
\draw[draw=none,fill=blue!20,thick] ($(ID-1-11.north west) + (0pt,2pt)$) rectangle
($(ID-1-15.south east) + (0pt,-2pt)$) node[midway] {$
\scalebox{0.75}{\FPic{7Sym12Sym45Sym67N}\FPic{7s24s367}\FPic{7ASym123ASym456N}\FPic{7s24s376}\FPic{7Sym12Sym45Sym67N}}
$};
\draw[stealth-,thick,red] (ID-1-10.north) |- ++(0,3mm)   -|
(ID-1-11.north) node[pos=.2,anchor=south,yshift=1mm] {\scriptsize absorb};
\draw[-stealth,thick,red] (ID-1-15.north) |- ++(0,3mm)   -|
(ID-1-16.north) node[pos=.2,anchor=south,yshift=1mm] {\scriptsize absorb};
\end{tikzpicture}\end{equation}
In particular, the ability
to factor out an operator $\tilde{O}$ within $O$ is encoded in the
requirement that the amputated tableau $\Ampc{\Theta}$ be
rectangular, as is discussed in section~\ref{sec:Propagating-transpositions}.

More generally, if $\Theta=\Tilde{\Theta}$ is a semi-standard irregular tableau, then
the following set of conditions on the amputated tableau will
determine whether certain symmetrizers can be propagated through
antisymmetrizers and vice versa:

\begin{theorem}[generalized propagation rules]\label{thm:PropagateCorollary}
  A form of the Propagation-Theorem~\ref{thm:PropagateSyms} holds also if $\Theta=\Tilde{\Theta}$
  is a semi-standard irregular tableau:
  \begin{equation}
    \label{eq:Propagate-general-Syms2}
    \mathbf{S}_{\Tilde{\Theta}} 
    \; \mathbf{A}_{\Tilde{\Theta}} 
    \; \mathbf{S}_{\Tilde{\Theta}\setminus\mathcal R}
    = 
    \mathbf{S}_{\Tilde{\Theta}\setminus\mathcal R} 
    \; \mathbf{A}_{\Tilde{\Theta}} 
    \; \mathbf{S}_{\Tilde{\Theta}}
    =
    \mathbf{S}_{\Tilde{\Theta}} 
    \; \mathbf{A}_{\Tilde{\Theta}} 
    \;\mathbf{S}_{\Tilde{\Theta}}
  \end{equation}
if all rows in $\Ampc{\Tilde{\Theta}}$ have equal lengths and 
  \begin{equation}
    \label{eq:Propagate-general-Syms3}
    \mathbf{A}_{\Tilde{\Theta}} 
    \; \mathbf{S}_{\Tilde{\Theta}} 
    \; \mathbf{A}_{\Tilde{\Theta}\setminus\mathcal C}
     = 
     \mathbf{A}_{\Tilde{\Theta}\setminus\mathcal C} 
     \; \mathbf{S}_{\Tilde{\Theta}} 
     \; \mathbf{A}_{\Tilde{\Theta}} 
    =
    \mathbf{A}_{\Tilde{\Theta}} 
    \; \mathbf{S}_{\Tilde{\Theta}} 
    \;\mathbf{A}_{\Tilde{\Theta}}
  \end{equation}
if all columns in $\Ampr{\Tilde{\Theta}}$ have equal
lengths. 

Requiring the amputated tableaux to have rows (resp. columns) of equal
lengths rather than them being rectangular allows for the fact that
$\cancel{\Tilde{\Theta}}$ (for $\Tilde{\Theta}$ being a semi-standard
irregular tableau) may contain disjoint pieces -- this cannot happen
for Young tableaux.\footnote{It should be noted that missing boxes
  within a row/column reduce its length, for example the first row of
  the tableau \scalebox{0.75}{$\begin{ytableau}
    1 & 2 & 3 \\
    4 & \none & 5
  \end{ytableau}$} has length $3$ but the second row only has length
$2$, thus corresponding to symmetrizers of length $3$ and $2$ respectively.}
\end{theorem}

The proof of Theorem~\ref{thm:PropagateSyms} only has to be altered in
minor ways to become a proof of the generalized
Propagation-Theorem~\ref{thm:PropagateCorollary}. These alterations
are given in section~\ref{sec:Generalized-Propagaration-Proof}.

As an example of Theorem~\ref{thm:PropagateCorollary}, consider the
semi-standard irregular tableau
\begin{equation}
\tilde{\Theta} = 
  \begin{ytableau}
    \none & 1 & 2 & 3 & 4 \\
    7 & 5 & 6 \\
   \none & 8 & 9
  \end{ytableau}
\qquad \text{with corresponding operator }
\Bar{Y}_{\Tilde{\Theta}} = \; \scalebox{0.75}{\FPic{9Sym1234Sym567Sym89N}\FPic{9s25697384}\FPic{9ASym123ASym456N}\FPic{9s24837965}}
\ ,
\end{equation}
\emph{cf.} eq.~\eqref{eq:Irreg-Tabl-Ops}. $\tilde{\Theta}$ is neither
a Young tableau nor does it uniquely specify
$\Bar{Y}_{\Tilde{\Theta}}$ as we could swap \ybox{3} and \ybox{4}
around and still obtain an appropriate tableau for
$\Bar{Y}_{\Tilde{\Theta}}$.  Let us now consider the operator
\begin{equation}
  O \; = \;
\underbrace{\underbrace{\scalebox{0.75}{\FPic{9Sym1234Sym567Sym89R}}
}_{\mathbf{S}_{\tilde{\Theta}}}
\underbrace{\scalebox{0.75}{\FPic{9s25697384}\FPic{9ASym123ASym456N}\FPic{9s24837965}}}_{\mathbf{A}_{\tilde{\Theta}}}}_{\Bar{Y}_{\Tilde{\Theta}}}
\underbrace{\scalebox{0.75}{\FPic{9Sym1234Sym567R}}}_{\mathbf{S}_{\tilde{\Theta}\setminus\mathcal{R}}}
\ ,
\end{equation}
where we would like to commute the symmetrizer $\bm{S}_{\mathcal{R}}=\bm{S}_{89}$ from the
left set $\mathbf{S}_{\tilde{\Theta}}$ to the right set
$\mathbf{S}_{\tilde{\Theta}\setminus\mathcal{R}}$. This symmetrizer
corresponds to the row $(8,9)$ in $\Tilde{\Theta}$,
\begin{equation}
\tilde{\Theta} = 
  \begin{ytableau}
    \none & 1 & 2 & 3 & 4 \\
    7 & 5 & 6 \\
   \none & *(magenta!20) 8 & *(magenta!20) 9
  \end{ytableau}
\ ;
\end{equation}
all rows in the column-amputated tableau of $\tilde{\Theta}$ according to the
row $(8,9)$ have equal lengths
\begin{equation}
  \cancel{\tilde{\Theta}}_{c}\left[(8,9)\right] = 
  \begin{ytableau}
    1 & 2\\
    5 & 6 \\
    *(magenta!20) 8 & *(magenta!20) 9
  \end{ytableau}
\ .
\end{equation}
Thus, by Theorem~\ref{thm:PropagateCorollary} we may propagate the symmetrizer
$\bm{S}_{89}$ from the left to the right, yielding
\begin{equation}
    O \; = \;
    \scalebox{0.75}{\FPic{9Sym1234Sym567Sym89N}\FPic{9s25697384}\FPic{9ASym123ASym456N}\FPic{9s24837965}\FPic{9Sym1234Sym567N}}
    \; = \;
    \scalebox{0.75}{\FPic{9Sym1234Sym567Sym89N}\FPic{9s25697384}\FPic{9ASym123ASym456N}\FPic{9s24837965}\FPic{9Sym1234Sym567Sym89N}}
    \; = \;
    \scalebox{0.75}{\FPic{9Sym1234Sym567N}\FPic{9s25697384}\FPic{9ASym123ASym456N}\FPic{9s24837965}\FPic{9Sym1234Sym567Sym89N}}
\ . 
\end{equation}
The reason why the propagation rule also works in this general case is
because an operator $\tilde{O}$ can be identified within $O$ in a
similar way as was done in~\eqref{eq:SimplificationEx1-TildeO}, see
section~\ref{sec:Generalized-Propagaration-Proof}.

\subsection{Proof of Theorem~\ref{thm:PropagateSyms} (propagation rules)}\label{sec:PropagateProofSection}

In this section, we provide a proof for eq.~\eqref{eq:PropagateSyms2}
of the  Propagation-Theorem~\ref{thm:PropagateSyms},
\begin{equation}
    O = \mathbf{S}_{\Theta} \; \mathbf{A}_{\Theta} \;
    \mathbf{S}_{\Theta\setminus\mathcal R}= \mathbf{S}_{\Theta} \; \mathbf{A}_{\Theta} \;
    \mathbf{S}_{\Theta}
    =
    \mathbf{S}_{\Theta\setminus\mathcal R} \; \mathbf{A}_{\Theta} \;\mathbf{S}_{\Theta}
\ .
\end{equation}
The proof of eq.~\eqref{eq:PropagateSyms3} (i.e. where the operator
$O$ is of the form
$O:=\mathbf{A}_{\Theta}\;\mathbf{S}_{\Theta}\;\mathbf{A}_{\Theta\setminus\mathcal{C}}$)
only changes in minor ways; these differences are discussed in
section~\ref{sec:PropagateSymsAnalysis}. 

The steps of the proof given in the present section can become rather
abstract; we therefore chose to accompany it with several schematic
drawings for clarification.

The strategy of this proof will be as follows: We start by
understanding what the conditions posed in
Theorem~\ref{thm:PropagateSyms} (in particular the requirement that
the amputated tableau be rectangular) imply for the
operator $O$. Then, we use a trick originally given
in~\cite{Keppeler:2013yla} to propagate the constituent permutations of
the symmetrizer $\bm{S}_{\mathcal{R}}$ through the set $\mathbf{A}_{\Theta}$
to the right of $O$; we recall that each symmetrizer is by
definition the sum of its constituent permutations,
\begin{equation}
  \bm{S}_{\mathcal{R}}=\frac{1}{\mathrm{length(\bm{S}_{\mathcal{R}})}!}\sum_{\rho}\rho
\ ,
\end{equation}
where $\rho$ are the constituent permutations of $\bm{S}_{\mathcal{R}}$, for
example
\begin{equation}
\label{eq:SymSumConstituents}
  \underbrace{\FPic{3Sym123SN}}_{\bm{S}_{123}} = 
\frac{1}{3!} 
\Bigl(
\,
 \underbrace{\FPic{3IdSN}}_{\mathrm{id}} 
+ \underbrace{\FPic{3s12SN}}_{(12)} 
+ \underbrace{\FPic{3s13SN}}_{(13)} 
+ \underbrace{\FPic{3s23SN}}_{(23)} 
+ \underbrace{\FPic{3s123SN}}_{(123)} 
+ \underbrace{\FPic{3s132SN}}_{(132)}
\,
\Bigr).
\end{equation}
The operators resulting from this propagation-process will then be summed up in the appropriate
manner\footnote{Similar to what was done in the example~\eqref{eq:HermitizeStep5}.}
to recombine to the symmetrizer $\bm{S}_{\mathcal{R}}$ on the right hand
side of $O$, yielding the desired
result. Let us thus begin:

\bigskip

Let $O:=\mathbf{S}_{\Theta}\mathbf{A}_{\Theta}\mathbf{S}_{\Theta\setminus\mathcal{R}}$ be an
operator as stated in the Propagation-Theorem~\ref{thm:PropagateSyms},
and let the sets $\mathbf{S}_{\Theta}$ and $\mathbf{S}_{\Theta\setminus\mathcal{R}}$ only differ by
one symmetrizer, $\bm{S}_{\mathcal{R}}$, with $\bm{S}_{\mathcal{R}}$ corresponding
to the row $\mathcal{R}$ in the Young tableau $\Theta$. We will
represent $O$ schematically as
\begin{equation}
  \label{eq:PropagateSymsProof1}
  O = \; \scalebox{0.75}{\FPic{mOpsO}}
\ ,
\end{equation}
where we used the fact that
$\mathbf{S}_{\Theta}=\mathbf{S}_{\Theta\setminus\mathcal{R}}\bm{S}_{\mathcal{R}}$, eq.~\eqref{eq:STheta-SR}.

\subsubsection{Unpacking the Theorem conditions:}\label{sec:Unpacking-Theorem-Conditions}

For the amputated tableau $\Ampc{\Theta}$ to be rectangular,
we clearly require all columns that overlap with the row $\mathcal{R}$
to have the same length. However, this is equivalent to saying that
every row in $\Theta$ other
than row $\mathcal{R}$ has to have length greater than or equal to
$\mathrm{length}(\mathcal{R})$: Suppose $\mathcal{R'}$ is a row in
$\Theta$ with
$\mathrm{length}(\mathcal{R'})<\mathrm{length}(\mathcal{R})$. Hence,
by definition of Young tableaux, the row $\mathcal{R'}$
is situated below the row $\mathcal{R}$. Furthermore, by the
left-alignedness of Young tableaux, this means that all the columns
that overlap with $\mathcal{R'}$ also overlap with $\mathcal{R}$; let
us denote this set of columns overlapping with the row $\mathcal{R}'$ by $C_{\mathcal{R'}}$. In addition,
there will be at least one column that overlaps with $\mathcal{R}$ but
does not overlap with $\mathcal{R'}$, since
$\mathrm{length}(\mathcal{R})>\mathrm{length}(\mathcal{R'})$; let us
denote this column by $\mathcal{C}$. Schematically, this situation can be
depicted as
\begin{equation}
  \label{eq:SchemPropYT}
\raisebox{-0.5\height}{\includegraphics{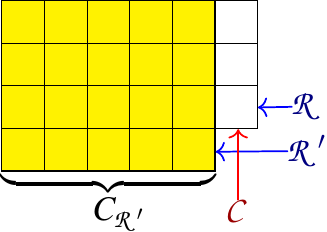}}
\ .
\end{equation}
It then follows by the top-alignedness of Young tableaux that
$\mathcal{C}$ is strictly shorter than the columns in the set
$C_{\mathcal{R'}}$, as is indicated in~\eqref{eq:SchemPropYT}. This
poses a contradiction, as we need all columns that overlap with
$\mathcal{R}$ to be of the same length for the tableau $\Ampc{\Theta}$
to be rectangular. Hence, there cannot be a row in $\Theta$ whose
length is strictly less than the length of $\mathcal{R}$.

Let $C_{\mathcal{R}}$ denote the set of columns overlapping with the
row $\mathcal{R}$. Since $\mathcal{R}$ is established to be (one of) the shortest row(s)
in $\Theta$, the top-alignedness and left-alignedness conditions of
Young tableaux imply thus that every other row in $\Theta$ also overlaps
with every column in $C_{\mathcal{R}}$.

In the language of symmetrizers, the discussion given above can be
formulated as:
\begin{enumerate}
\item\label{itm:ProofCondition1} $\bm{S}_{\mathcal{R}}$ (corresponding to
  the row $\mathcal{R}$ of $\Theta$) is the shortest symmetrizer in the set
  $\mathbf{S}_{\Theta}$.
\item\label{itm:ProofCondition2} Each leg of $\bm{S}_{\mathcal{R}}$ enters an antisymmetrizer in
  $\mathbf{A}_{\Theta}$ of equal length; let us denote this subset of
  antisymmetrizer by $\mathbf{A'}_{\bm{S}_{\mathcal{R}}}$ (this set of
  antisymmetrizers correspond to the set of columns $C_{\mathcal{R}}$).
\item\label{itm:ProofCondition3} Each symmetrizer in
  $\mathbf{S}_{\Theta}$ has one common leg with each antisymmetrizer
  in $\mathbf{A'}_{\bm{S}_{\mathcal{R}}}$ (since each row in $\Theta$
  overlaps with each column in $C_{\mathcal{R}}$).
\item\label{itm:ProofCondition4} Since, by the assumption of the
  Propagation-Theorem, $\mathbf{S}_{\Theta\setminus\mathcal{R}}$ and
  $\mathbf{S}_{\Theta}$ only differ by the symmetrizer
  $\bm{S}_{\mathcal{R}}$, this means that each symmetrizer in the set
  $\mathbf{S}_{\Theta\setminus\mathcal{R}}$ has a common leg with each
  antisymmetrizer in the set $\mathbf{A'}_{\bm{S}_{\mathcal{R}}}$.
\end{enumerate}

\subsubsection{Strategy of the proof:}\label{sec:Propagation-Proof-Strategy}

In this proof, we will use the fact that the symmetrizer
$\bm{S}_{\mathcal{R}}$ by definition is the sum of all permutations of the
legs over which $\bm{S}_{\mathcal{R}}$ is drawn. If $\bm{S}_{\mathcal{R}}$ has
length $k$, then this sum will consist of $k!$ terms, and there will
be a constant prefactor $\nicefrac{1}{k!}$; this was exemplified
in~\eqref{eq:SymSumConstituents}. In particular, if $\lambda$ is a
particular permutation in the series expansion of $\bm{S}_{\mathcal{R}}$,
then we will show that $O = O^{\lambda}$, where $O^{\lambda}$ is
defined to be the operator $O$ with the permutation $\lambda$ added on
the right side in the place where $\bm{S}_{\mathcal{R}}$ would
be; schematically drawn, we will show that
\begin{equation}
  \label{eq:PropagateSymsProof2}
  O = \;
\scalebox{0.75}{\FPic{mOpsO}} 
\; \stackrel{?}{=} \;
\scalebox{0.75}{\FPic{mOpsOLambda}} 
\; =: O^{\lambda}
\ .
\end{equation}
Since the constituent permutations of a
  symmetrizer over a subset of factors in $\Pow{n}$ form a sub-group of
  $S_n$~\cite{Tung:1985na}, it immediately follows that every
  constituent permutation of $\bm{S}_{\mathcal{R}}$ can be written as
  a product of constituent transpositions of
  $\bm{S}_{\mathcal{R}}$.\footnote{A proof that any permutation in $S_n$ can be written as
  the product of transpositions can be found in~\cite{Artin:2011} and
  other standard textbooks.} It thus suffices to show
that~\eqref{eq:PropagateSymsProof2} holds for $\lambda$ being a
constituent transposition of $\bm{S}_{\mathcal{R}}$ (i.e. that we may
propagate a transposition from the left symmetrizer
$\bm{S}_{\mathcal{R}}$ to the right), as then any other permutation
can be produced by the successive propagation of transpositions.

\subsubsection{Propagating transpositions:}\label{sec:Propagating-transpositions}

The technique used to permute transpositions through the set
of antisymmetrizers, as described in the previous paragraph, was
inspired by an example presented in the Appendix of~\cite{Keppeler:2013yla}.

\begin{figure}[ht]
\centering
\scalebox{0.5}{\includegraphics{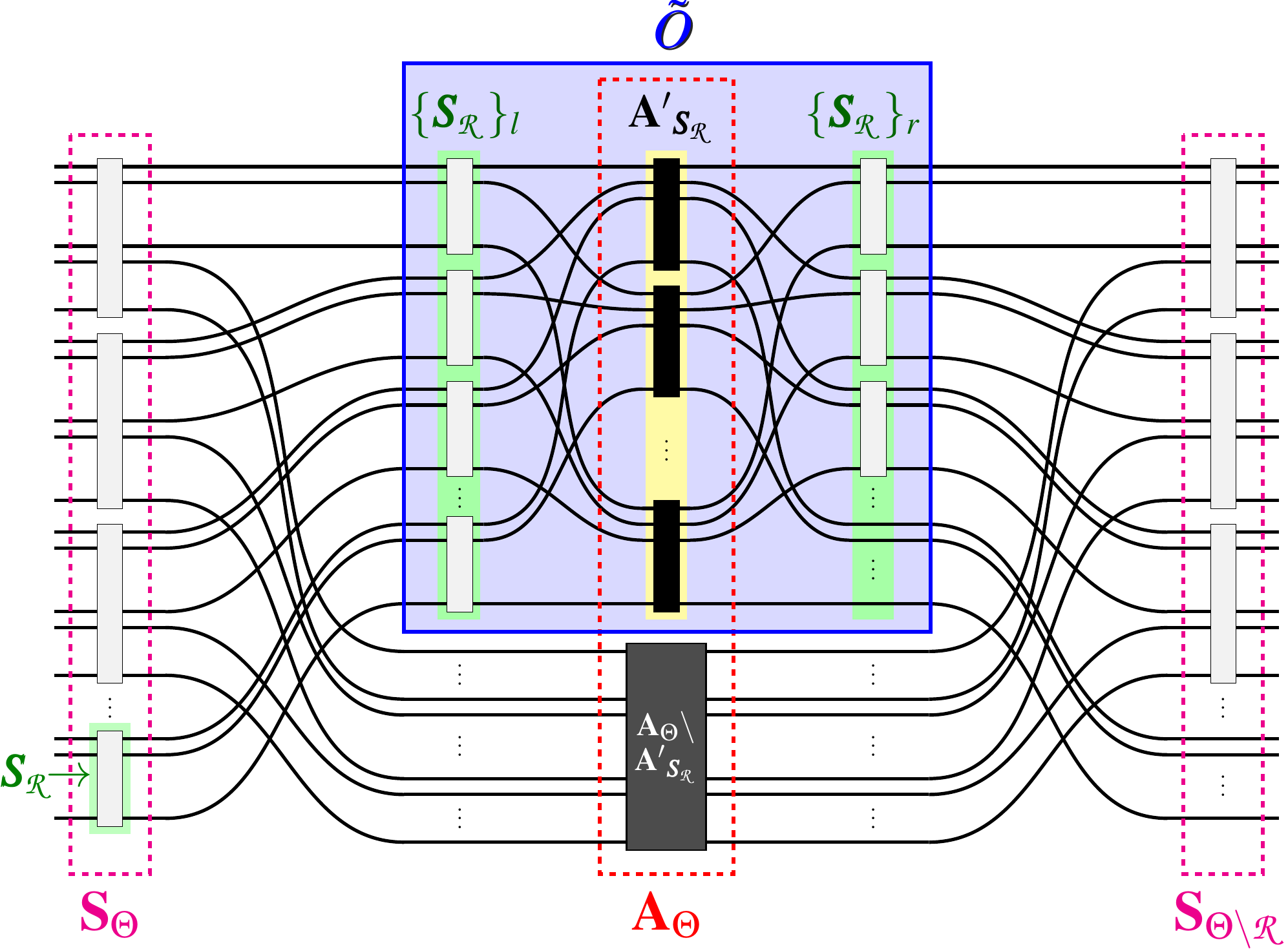}}\caption{This diagram schematically depicts the operator $O$,~\eqref{eq:PropagateSymsProof1}, with a symmetrizer $\bm{S}_{\mathcal{R}}$
  factored out of each symmetrizer in $\mathbf{S}_{\Theta}$ and
  $\mathbf{S}_{\Theta\setminus\mathcal{R}}$. The left set of $\bm{S}_{\mathcal{R}}$'s will be denoted
  by $\lbrace\bm{S}_{\mathcal{R}}\rbrace_l$, and the right set of
  $\bm{S}_{\mathcal{R}}$'s by $\lbrace\bm{S}_{\mathcal{R}}\rbrace_r$. In this proof, we
  will focus on the part of the operator that is highlighted in
 blue. This part will be denoted by $\tilde{O}$.}
\label{fig:PropagateSyms1}
\end{figure}

Suppose the set $\mathbf{A'}_{\bm{S}_{\mathcal{R}}}$ introduced in
condition~\ref{itm:ProofCondition2} of the previous discussion
contains $n$ antisymmetrizers. Then, by
observations~\ref{itm:ProofCondition1}-\ref{itm:ProofCondition4}, the
length of $\bm{S}_{\mathcal{R}}$ will be exactly $n$, and each other
symmetrizer in $\mathbf{S}_{\Theta}$ (and thus also each symmetrizer
in $\mathbf{S}_{\Theta\setminus\mathcal{R}}$) will have length at
least $n$. We may then factor ``the symmetrizer
$\bm{S}_{\mathcal{R}}$'' (i.e. a symmetrizer of length $n$) out of
each symmetrizer in the sets $\mathbf{S}_{\Theta}$ and
$\mathbf{S}_{\Theta\setminus\mathcal{R}}$,
\begin{equation}
  \raisebox{-60pt}{\scalebox{0.32}{\includegraphics{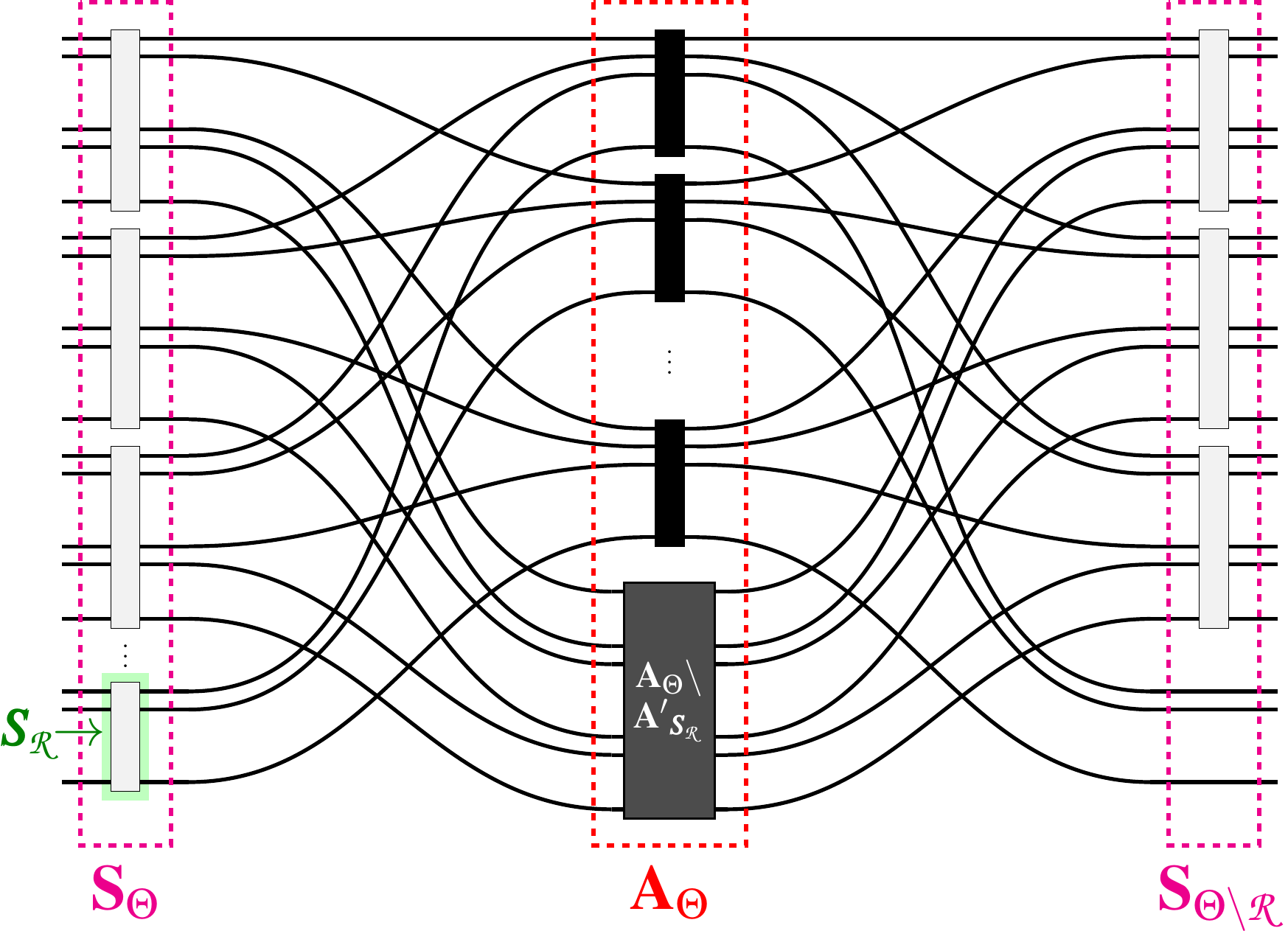}}}
\hspace{0.7cm} \xrightarrow{\text{factor $\bm{S}_{\mathcal{R}}$}} \hspace{0.7cm}
\raisebox{-60pt}{\scalebox{0.32}{\includegraphics{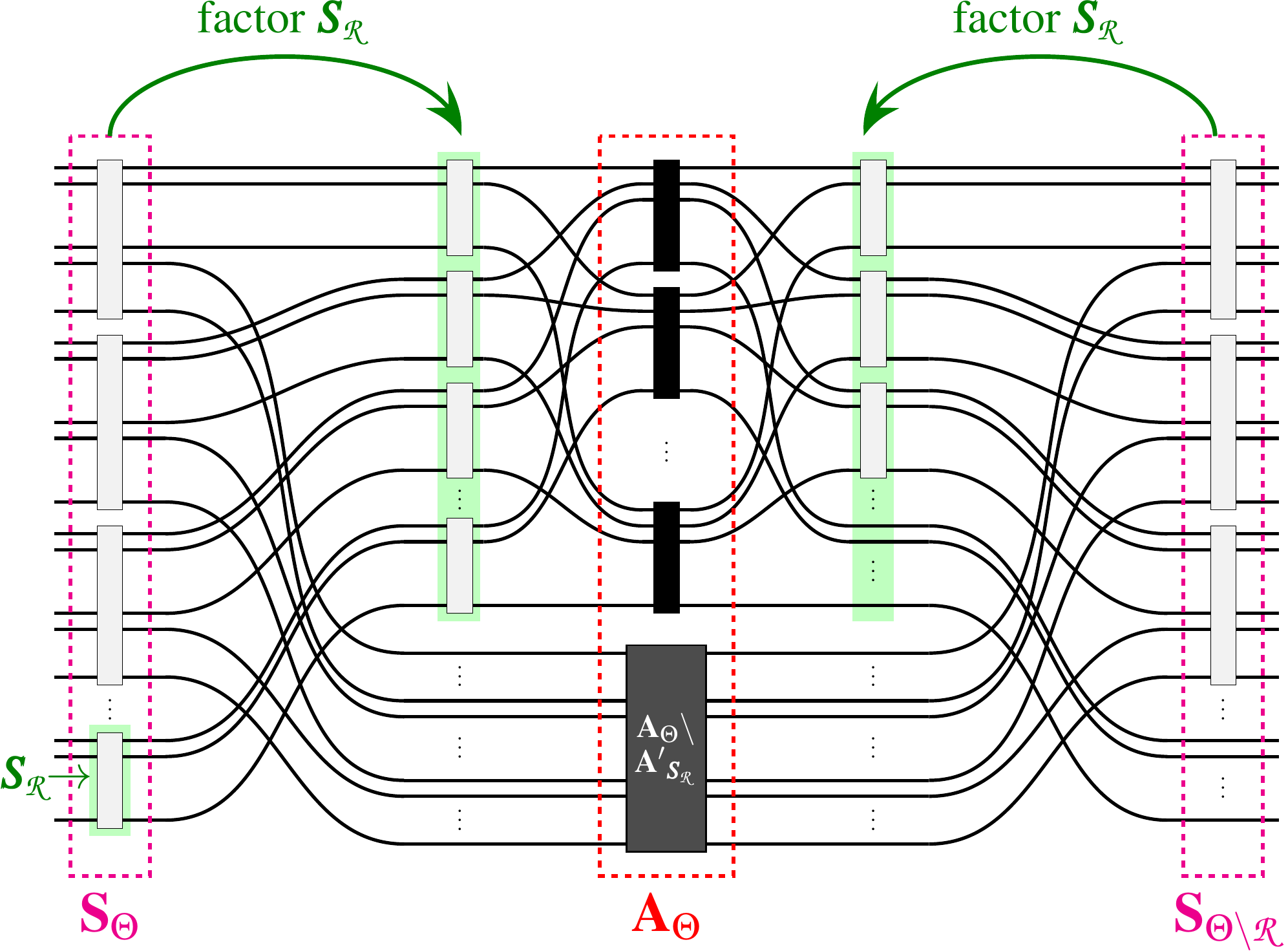}}}
\ ,
\end{equation}
where we lumped together the antisymmetrizers
$\mathbf{A'}_{\bm{S}_{\mathcal{R}}}$ and the rest ($\mathbf{A}_{\Theta}\setminus\mathbf{A'}_{\bm{S}_{\mathcal{R}}}$). We will denote the left set of
$\bm{S}_{\mathcal{R}}$'s (which were factored out of $\mathbf{S}_{\Theta}$)
by $\lbrace\bm{S}_{\mathcal{R}}\rbrace_l$, and the right set of
$\bm{S}_{\mathcal{R}}$'s (which were factored out of $\mathbf{S}_{\Theta\setminus\mathcal{R}}$) by
$\lbrace\bm{S}_{\mathcal{R}}\rbrace_r$, see
Figure~\ref{fig:PropagateSyms1}. From now onwards, we will focus the
part of the operator $O$ that is highlighted blue in
Figure~\ref{fig:PropagateSyms1}.

\paragraph{The significance of the operator $\bm{\tilde{O}}$ in Fig.~\ref{fig:PropagateSyms1}:} The left
part of $\tilde{O}$, namely
$\lbrace\bm{S}_{\mathcal{R}}\rbrace_l\cdot\mathbf{A'}_{\bm{S}_{\mathcal{R}}}$,
by itself corresponds to a rectangular tableau, as each symmetrizer
has the same length and each antisymmetrizer has the same
length. This will be important, since we will need $\tilde{O}$ to stay
unchanged under a swap of any pair of antisymmetrizers in
$\mathbf{A'}_{\bm{S}_{\mathcal{R}}}$ in order to commute the
constituent permutations of $\bm{S}_{\mathcal{R}}$ (in analogy to what
was done in example~\eqref{eq:HermitizeStep3} -- this will become
evident below). Note that $\tilde{O}$ would not stay unchanged under
such a swap if the antisymmetrizers in
$\mathbf{A'}_{\bm{S}_{\mathcal{R}}}$ had different length's and would
thus be distinguishable. In particular, the operator $\tilde{O}$
corresponds to the amputated tableau $\Ampc{\Theta}$, which is indeed
rectangular by requirement of the
Propagation-Theorem~\ref{thm:PropagateSyms}. This requirement
therefore translates into the ability of finding an operator
$\tilde{O}$ within the operator $O$, thus allowing the necessary
propagation of permutations.

\bigskip

\indent Suppose that
$\mathbf{S}_{\Theta}$ contains exactly $m$ symmetrizers (hence
$\mathbf{S}_{\Theta\setminus\mathcal{R}}$ contains $(m-1)$ symmetrizers). Then
$\lbrace\bm{S}_{\mathcal{R}}\rbrace_l$ will contain $m$ symmetrizers and
$\lbrace\bm{S}_{\mathcal{R}}\rbrace_r$ contains $(m-1)$ symmetrizers.

Furthermore, since each symmetrizer in
$\lbrace\bm{S}_{\mathcal{R}}\rbrace_l$ has a common leg with each of
the $n$ antisymmetrizer in $\mathbf{A'}_{\bm{S}_{\mathcal{R}}}$, we
may choose the $k^{th}$ leg exiting each symmetrizer in
$\lbrace\bm{S}_{\mathcal{R}}\rbrace_l$ to enter the $k^{th}$
antisymmetrizer in $\mathbf{A'}_{\bm{S}_{\mathcal{R}}}$.\footnote{We
  may always choose to order index legs this way, since, within a
  symmetrizer, we may re-order index lines at will without changing
  the symmetrizer.} We may schematically draw this, as
\begin{equation}
\label{eq:OTilde-ASyms}
\scalebox{0.75}{\FPic{mSymnNlabelF}\FPic{mnPermN}\FPic{nASymmNlabelF}}
\ ;
\end{equation}
In~\eqref{eq:OTilde-ASyms}, we have labelled the index lines for clarity; from
Figure~\ref{fig:PropagateSyms1} it however should be noted that the
$i^{th}$ index in the above graphic is not necessarily the $i^{th}$
index line in the operator $O$. The part of the operator
$O$ highlighted in blue in Figure~\ref{fig:PropagateSyms1}, operator $\tilde{O}$, can then be represented as
\begin{equation}
\label{eq:OTilde-Label-ASym12}
 \mbox{\Large $\tilde{O}$ \; = \;}
 \scalebox{0.75}{$\underbrace{\FPic{mSymnNlabelF}}_{\mbox{\normalsize
       $\lbrace\bm{S}_{\mathcal{R}}\rbrace_l$}}\FPic{mnPermN}\underbrace{\FPic{nASymmN12F}}_{\mbox{\normalsize
       $\mathbf{A'}_{\bm{S}_{\mathcal{R}}}$}}\FPic{mnPermInvN}\underbrace{\FPic{m-1_SymnNF}}_{\mbox{\normalsize
       $\lbrace\bm{S}_{\mathcal{R}}\rbrace_r$}}$}
\ ,
\end{equation}
where the last symmetrizer in the set
$\lbrace\bm{S}_{\mathcal{R}}\rbrace_l$ is the symmetrizer
$\bm{S}_{\mathcal{R}}$ which we eventually wish to commute through to
the right. In~\eqref{eq:OTilde-Label-ASym12}, we labeled the first and
the second antisymmetrizer of the set
$\mathbf{A'}_{\bm{S}_{\mathcal{R}}}$ by \scalebox{0.85}{\color{red}
  \circled{1}} and \scalebox{0.85}{\color{red} \circled{2}}
respectively for future reference.

As previously stated, we strive to commute constituent transpositions
$(ij)$ of the symmetrizer
$\bm{S}_{\mathcal{R}}\in\lbrace\bm{S}_{\mathcal{R}}\rbrace_l$ through the set
of anti-symmetriers $\mathbf{A'}_{\bm{S}_{\mathcal{R}}}$ to the right set
$\lbrace\bm{S}_{\mathcal{R}}\rbrace_r$. We achieve this goal in the
following way: We first factor the transposition $(ij)$ out of
\emph{each} symmetrizer in $\lbrace\bm{S}_{\mathcal{R}}\rbrace_l$. By
doing so, the $i^{th}$ leg of each symmetrizer now enters the $j^{th}$
antisymmetrizer and vice versa; all the other legs stay unchanged. We
may now ``remedy'' this change by swapping the $i^{th}$ and $j^{th}$
antisymmetrizer, similar to what we did in
example~\eqref{eq:HermitizeStep3}. For instance, if $i=1$ and $j=2$, we factor the 
transposition $(12)$ out of each of the symmetrizers of
$\lbrace\bm{S}_{\mathcal{R}}\rbrace_l$,
\begin{equation}
 \mbox{\Large $\tilde{O}$ \; = \;}
\scalebox{0.75}{\FPic{mSymnNlabelF}{\setlength{\fboxsep}{1pt}\colorbox{red!20}{\FPic{mns12allNlabel}}}\FPic{mnPermN}\FPic{nASymmN12F}\FPic{mnPermInvN}\FPic{m-1_SymnNF}}
\ ,
\end{equation}
and then swap the first and second antisymmetrizer, which are marked
as \scalebox{0.85}{\color{red} \circled{1}} and \scalebox{0.85}{\color{red} \circled{2}}
respectively.  The key observation to make is that the antisymmetrizers
\scalebox{0.85}{\color{red} \circled{1}} and \scalebox{0.85}{\color{red} \circled{2}} would be
indistinguishable if it weren't for the labeling. Thus, the set $\mathbf{A'}_{\bm{S}_{\mathcal{R}}}$ remains unchanged even when the
swap between antisymmetrizers $i$ (\scalebox{0.85}{\color{red}
  \circled{1}}) and $j$ (\scalebox{0.85}{\color{red} \circled{2}}) is carried out. This trick of swapping identical
antisymmetrizers was initially used by KS in an example in the appendix of~\cite{Keppeler:2013yla}.

After we swapped the two antisymmetrizers, the $i^{th}$ leg of each
symmetrizer in $\lbrace\bm{S}_{\mathcal{R}}\rbrace_l$ once again
enters the $i^{th}$ antisymmetrizer and same is true for the $j^{th}$
leg. However, now the legs exiting the $i^{th}$ antisymmetrizer
$\mathbf{A'}_{\bm{S}_{\mathcal{R}}}$ enter the symmetrizers in
$\lbrace\bm{S}_{\mathcal{R}}\rbrace_r$ in the $j^{th}$ position, and
the legs exiting the $j^{th}$ antisymmetrizer enter the symmetrizers
in $\lbrace\bm{S}_{\mathcal{R}}\rbrace_r$ in the $i^{th}$ position.
Thus, we have effectively commuted the transpositions $(ij)$ past the
set $\mathbf{A'}_{\bm{S}_{\mathcal{R}}}$,
\begin{equation}
 \mbox{\Large $\tilde{O}$ \; = \;}
\scalebox{0.75}{\FPic{mSymnNlabelF}\FPic{mnPermN}\FPic{nASymmN21F}\FPic{mnPermInvN}{\setlength{\fboxsep}{1pt}\colorbox{red!20}{\FPic{mns12allNlabel}}}\FPic{m-1_SymnNF}}
\ .
\end{equation}

\noindent All but one of the propagated transpositions $(ij)$ can then
be absorbed into the symmetrizers of the set
$\lbrace\bm{S}_{\mathcal{R}}\rbrace_r$. The bottom transposition will remain, as
there is no symmetrizer in the set
$\lbrace\bm{S}_{\mathcal{R}}\rbrace_r$ to
absorb this transposition,
\begin{equation}
 \mbox{\Large $\tilde{O}$ \; = \;}
\scalebox{0.75}{\FPic{mSymnNlabelF}\FPic{mnPermN}\FPic{nASymmN21F}\FPic{mnPermInvN}{\setlength{\fboxsep}{1pt}\colorbox{red!20}{\FPic{m-1_Symns12lastNF}}}}
\ .
\end{equation}
We then re-absorb the sets $\lbrace\bm{S}_{\mathcal{R}}\rbrace_l$ and
$\lbrace\bm{S}_{\mathcal{R}}\rbrace_l$ into $\mathbf{S}_{\Theta}$ and
$\mathbf{S}_{\Theta\setminus\mathcal{R}}$ respectively. This clearly leaves the transposition
$(ij)$ un-absorbed. Thus, we have shown that
\begin{equation}
  \label{eq:PropagateSymsProof3}
  O = \; \scalebox{0.75}{\FPic{mOpsO}} \; = \;
  \scalebox{0.75}{\FPic{mOpsOLambda}} \; = O^{\lambda}
\end{equation}
for $\lambda=(ij)$ being a transposition. We can repeat the above
procedure with any constituent tranposition of $\bm{S}_{\mathcal{R}}$.

If $\lambda$ is a constituent permutation (not
necessarily transposition) of $\bm{S}_{\mathcal{R}}$, we can also
propagate $\lambda$ to the right hand side, since any
such permutation $\lambda$ can be written as a product of constituent
transpositions; propagating the permutation $\lambda$ then
corresponds to successively propagating the constituent transpositions through to
the right, yielding
\begin{equation}
  \mbox{\Large $\tilde{O} \; = \; $} \scalebox{0.75}{\FPic{mSymnNlabelF}\FPic{mnPermN}\FPic{nASymmNF}\FPic{mnPermInvN}\FPic{m-1_SymnlambdalastNF}}
\end{equation}
for any constituent permutation $\lambda$ of $\bm{S}_{\mathcal{R}}$.

In order to obtain the missing symmetrizer on the right, it remains to
add up all the terms $O^{\lambda}$ -- since $\bm{S}_{\mathcal{R}}$ is
assumed to have length $n$ there will be exactly $n!$ such terms. By relation
\eqref{eq:PropagateSymsProof3}, we know that each of these terms is
equal to $O$, yielding the following sum,
\begin{equation}
  \frac{1}{n!} \displaystyle \sum_1^{n!} \left(\underbrace{\scalebox{0.75}{\FPic{mOpsO}}}_{O}\right) \; = \;
    \frac{1}{n!} \displaystyle \sum_{\lambda \in S_n} \left(\underbrace{\scalebox{0.75}{\FPic{mOpsOLambda}}}_{O^{\lambda}}\right).
\end{equation}
The left hand side of the above equation merely becomes $\frac{n!}{n!}
O = O$. The right hand side yields the desired
symmetrizer,\footnote{This was already exhibited in example~\eqref{eq:HermitizeStep5}.} such that
\begin{equation}
  \label{eq:RectangularProjector7}
O = \mathbf{S}_{\Theta} \; \mathbf{A}_{\Theta} \; \mathbf{S}_{\Theta}
= \; \scalebox{0.75}{\FPic{mOpsOHermitean}}
\ ,
\end{equation}
where we used the fact that  $\mathbf{S}_{\Theta}=\mathbf{S}_{\Theta\setminus\mathcal R}\bm{S}_{\mathcal R}=\bm{S}_{\mathcal R}\mathbf{S}_{\Theta\setminus\mathcal R}$ 
by assumption of Theorem~\ref{thm:PropagateSyms} (\emph{c.f.} eq.~\eqref{eq:STheta-SR}).  In particular,
using the fact that $O$ as given in~\eqref{eq:RectangularProjector7}
is clearly Hermitian, $O^{\dagger}=O$,\footnote{By the Hermiticity of
  (sets of) (anti-) symmetrizer, see~\eqref{eq:ASymsHC}.} we find that
\begin{equation}
O =\; \scalebox{0.75}{\FPic{mOpsO}} \; = \;
\scalebox{0.75}{\FPic{mOpsOHermitean}} \; = \;
\scalebox{0.75}{\FPic{mOpsOHC}} \; = O^{\dagger},
\end{equation}
as required.

\subsubsection{Propagating antisymmetrizers:}\label{sec:PropagateSymsAnalysis}

The proof of the Propagation-Theorem~\ref{thm:PropagateSyms} for
an operator $Q$ of the form
$Q:=\mathbf{A}_{\Theta}\mathbf{S}_{\Theta}\mathbf{A}_{\Theta\setminus\mathcal{C}}$ is very
similar to the proof given for the operator $O$, however there are
some differences on which we
wish to comment here: If we want to propagate an antisymmetrizer
$\bm{A}_{\mathcal{C}}$ corresponding to a column $\mathcal{C}$ in $\Theta$
from $\mathbf{A}_{\Theta}$ to $\mathbf{A}_{\Theta\setminus\mathcal{C}}$, we first check that the
amputated tableau $\Ampr{\Theta}$ is rectangular. If so, we are able
to isolate an operator $\tilde{Q}$ within $Q$ in the same way as we
did for $\tilde{O}$ within $O$, see Figure~\ref{fig:PropagateSyms1},
where
\begin{equation}
  \tilde{Q}
:=
\lbrace\bm{A}_{\mathcal{C}}\rbrace_l \;
\mathbf{S'}_{\bm{A}_{\mathcal{C}}} \;
\lbrace\bm{A}_{\mathcal{C}}\rbrace_r
\ .
\end{equation}
When we propagate a transposition $(ij)$ from the left to the
right of $\tilde{Q}$, we need to tread with care as this will induce a
factor of $(-1)$; this factor however will be vital in the
recombination-process where we recreate the antisymmetrizer
$\bm{A}_{\mathcal{C}}$ by summing constituent permutations: Suppose
the set $\lbrace\bm{A}_{\mathcal{C}}\rbrace_l$ contains $m$
antisymmetrizers, then the set $\lbrace\bm{A}_{\mathcal{C}}\rbrace_r$
contains $(m-1)$ antisymmetrizers. If we now factor a transposition
$(ij)$ out of each antisymmetrizer in
$\lbrace\bm{A}_{\mathcal{C}}\rbrace_l$ on the left of $\tilde{Q}$, we
obtain a factor of $(-1)^m$. Swapping the $i^{th}$ and $j^{th}$
symmetrizers will not induce an extra minus-sign, but absorbing the
transpositions into the antisymmetrizers in the set
$\lbrace\bm{A}_{\mathcal{C}}\rbrace_r$ will produce an extra factor of
$(-1)^{m-1}$. Thus, for each transposition we commute through, we
obtain a factor of $(-1)^{2m-1}=-1$, which is the signature of a
transposition. In particular, each permutation $\lambda$ (consisting
of a product of transpositions) will induce a pre-factor of
$\mathrm{sign}(\lambda)$ when commuted through, yielding
\begin{equation}
\label{eq:PropagationAnalysis1}
  \tilde{Q} = \mathrm{sign}(\lambda) \tilde{Q}^{\lambda}.
\end{equation}
However, since an antisymmetrizer is by definition the sum of its
constituent permutations \emph{weighted by their signatures}, for
example,
\begin{equation}
  \FPic{3ASym123SN} = \frac{1}{3!} \left( \FPic{3IdSN} - \FPic{3s12SN}
    - \FPic{3s13SN} - \FPic{3s23SN} + \FPic{3s123SN} + \FPic{3s132SN}
  \right)
\ ,
\end{equation}
equation~\eqref{eq:PropagationAnalysis1} is exactly what we need in
order to be able to reconstruct the antisymmetrizer
$\bm{A}_{\mathcal{C}}$ on the right of the operator $\tilde{Q}$ by
summing up the terms $\mathrm{sign}(\lambda) \tilde{Q}^{\lambda}$.
Re-absorbing $\lbrace\bm{A}_{\mathcal{C}}\rbrace_l$ into
$\mathbf{A}_{\Theta}$ and $\lbrace\bm{A}_{\mathcal{C}}\rbrace_r$ into
$\mathbf{A}_{\Theta\setminus\mathcal{C}}$ yields the desired
eq.~\eqref{eq:PropagateSyms3}. \qed

\subsection{Proof of Theorem~\ref{thm:PropagateCorollary} (generalized
  propagation rules)}\label{sec:Generalized-Propagaration-Proof}

Let $\Tilde{\Theta}$ be a semi-standard irregular tableau and let $O$
be an operator of the form
\begin{equation}
  O \; = \;
    \mathbf{S}_{\Tilde{\Theta}} 
    \; \mathbf{A}_{\Tilde{\Theta}} 
    \; \mathbf{S}_{\Tilde{\Theta}\setminus\mathcal R}
    \ ,
\end{equation}
where $\mathcal{R}$ denotes a particular row in $\Tilde{\Theta}$.
Having gone through the proof of Theorem~\ref{thm:PropagateSyms}, it is
clear that the symmetrizer $\bm{S}_{\mathcal{R}}$ can be propagated
through the set $\mathbf{A}_{\Tilde{\Theta}}$ if the following
conditions are met:
\begin{enumerate}
\item Each leg in
  $\bm{S}_{\mathcal{R}}$ needs to enter an antisymmetrizer of the same
  length to perform the swapping procedure described in
  section~\ref{sec:Propagating-transpositions}. Call the set of
  antisymmetrizers sharing legs with $\bm{S}_{\mathcal{R}}$
  $\mathbf{A'}_{\bm{S}_{\mathcal{R}}}$. If $\bm{S}_{\mathcal{R}}$ has
  length $n$, then $\mathbf{A'}_{\bm{S}_{\mathcal{R}}}$ contains
  exactly $n$ antisymmetrizers.
\item The remaining symmetrizers in $O$ may
  have
  \begin{itemize}
  \item $0$ legs in common with the antisymmetrizers in
    $\mathbf{A'}_{\bm{S}_{\mathcal{R}}}$, i.e. they are not affected
    by swapping the antisymmetrizers in
    $\mathbf{A'}_{\bm{S}_{\mathcal{R}}}$ and thus do not contribute to
    $\Tilde{O}$ (\emph{c.f.}
    section~\ref{sec:Propagating-transpositions}), or
  \item $n$ legs in common with the antisymmetrizers in
    $\mathbf{A'}_{\bm{S}_{\mathcal{R}}}$ (one with each
    antisymmetrizer) in order to be able to absorb any permutation
    arising from swapping the antisymmetrizers in $\mathbf{A'}_{\bm{S}_{\mathcal{R}}}$.
  \end{itemize}
\end{enumerate}
The requirement that $\Ampc{\Tilde{\Theta}}$ only
contains rows of equal length ensures that the above conditions are
met:

Differently to Young tableaux, if $\Tilde{\Theta}$ is a semi-standard
irregular tableau then removing columns in order to form
$\Ampc{\Tilde{\Theta}}$ may remove whole rows in the process, for
example
\begin{equation}
\Tilde{\Theta} = \;
  \begin{ytableau}
    \none & \none & 5 & 1 \\
    7 & 2 & 3 & 6 \\
    *(green) 4 & *(green) 8
  \end{ytableau}
\quad \longrightarrow \quad
\cancel{\Tilde{\Theta}}_c[(4,8)] \;
  \begin{ytableau}
    7 & 2 \\
    *(green) 4 & *(green) 8
  \end{ytableau}
\ ,
\end{equation}
where the row $(5,1)$ was removed. However, the
symmetrizers corresponding to such rows have no common legs with the
antisymmetrizers in $\mathbf{A'}_{\bm{S}_{\mathcal{R}}}$, as is
evident from the example ($\bm{S}_{51}$ shares no legs with
$\bm{A}_{74}$ and $\bm{A}_{28}$). This is also due to the fact that
$\Tilde{\Theta}$ is semi-standard, thus not allowing any of its
entries to occur more than once.

Hence, the only symmetrizers that share index legs with the
antisymmetrizers in $\mathbf{A'}_{\bm{S}_{\mathcal{R}}}$ are those
corresponding to the rows of $\Tilde{\Theta}$ which have not been
fully deleted (although maybe in part) in $\Ampc{\Tilde{\Theta}}$. Let
us denote the set of these symmetrizers by
$\mathbf{S'}_{\bm{S}_{\mathcal{R}}}$.  The requirement that each row
in $\Ampc{\Tilde{\Theta}}$ has the same length ensures each
symmetrizer in $\mathbf{S'}_{\bm{S}_{\mathcal{R}}}$ shares exactly one
leg with each antisymmetrizer in $\mathbf{A'}_{\bm{S}_{\mathcal{R}}}$
and thus has length $\geq n$
(this was already argued in
section~\ref{sec:Unpacking-Theorem-Conditions}).

Thus, all conditions required to perform the propagating procedure
already explained in sections~\ref{sec:Propagation-Proof-Strategy} and
~\ref{sec:Propagating-transpositions} are met, allowing us to
propagate $\bm{S}_{\mathcal{R}}$ through
$\mathbf{A'}_{\bm{S}_{\mathcal{R}}}$ at will,
\begin{equation}
  O \; = \;
    \mathbf{S}_{\Tilde{\Theta}} 
    \; \mathbf{A}_{\Tilde{\Theta}} 
    \; \mathbf{S}_{\Tilde{\Theta}\setminus\mathcal R}
\; = \;
    \mathbf{S}_{\Tilde{\Theta}} 
    \; \mathbf{A}_{\Tilde{\Theta}} 
    \; \mathbf{S}_{\Tilde{\Theta}}
\; = \;
    \mathbf{S}_{\Tilde{\Theta}\setminus\mathcal R} 
    \; \mathbf{A}_{\Tilde{\Theta}} 
    \; \mathbf{S}_{\Tilde{\Theta}}
    \ .
\end{equation}
The proof for symmetrizers and antisymmetrizers exchanged follows
similar steps and is thus left as an exercise to the reader. \qed

\section{Conclusion}\label{sec:Conclusion}

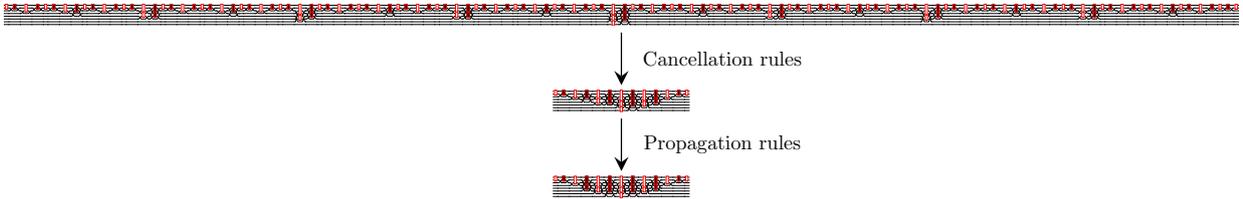
\begin{figure}[H]
  \begin{center}
\resizebox{\textwidth}{!}{\begin{tikzpicture}[every node/.style={inner sep=1pt, outer sep=0pt}]
\node (KS) {\diagram[height=.15cm]{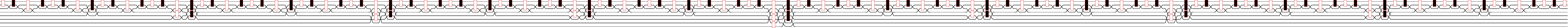}};
\node (short) at ($(KS) +(0,-0.6cm)$)
      {\diagram[height=.15cm]{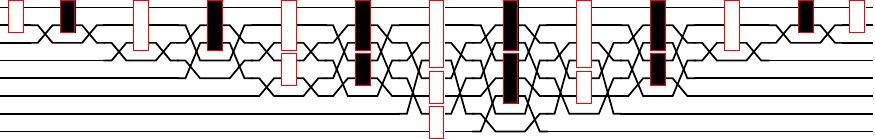}};
\node (MOLD) at ($(short)
      +(0,-0.6cm)$)  {\diagram[height=.15cm]{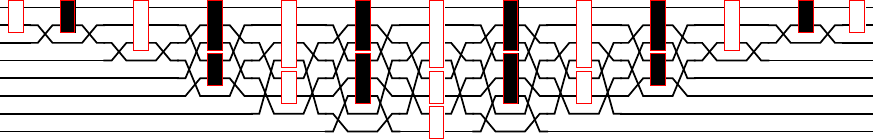}};
\draw[-{stealth}, line width=0.25pt] (KS) to (short);
\draw[-{stealth}, line width=0.25pt] (short) to (MOLD);
\node[scale=0.4] (Cancel) at ($(KS) +(0.7,-0.3cm)$) {Cancellation
  rules};
\node[scale=0.4] (Propagate) at ($(short) +(0.7,-0.3cm)$) {Propagation rules};
\end{tikzpicture}
}
  \end{center}
\caption{For a size comparison, this figure shows the birdtrack
  arising from the iterative KS-construction in the top line, the much
  shortened version arising from the cancellation rule in the second
  line, and the explicitly symmetric version achieved via the
  propagation rules in the third line.}
\label{fig:MOLDAdvantage}
\end{figure}

We have established two classes of rules which allow us to simplify and
manipulate birdtrack operators. The first such class are the
\emph{cancellation rules}, Theorem~\ref{thm:CancelWedgedYoung} and Corollary~\ref{thm:Cancel-Ops}, allowing us
to shorten the birdtrack expression of an operator. The simplification
reached in this process is often very significant, as is exemplified
in Figure~\ref{fig:MOLDAdvantage}. Shorter expressions of operators
are desirable, as they are more practical to work with in that they
allow for faster automated computation. Furthermore, short expressions
offer a visual assessment of their action, making them more intuitive
to work with.

The second class of rules are \emph{propagation
  rules}, Theorems~\ref{thm:PropagateSyms} and~\ref{thm:PropagateCorollary}. Their use lies in the
ability to make Hermitian birdtrack operators explicitly symmetric,
thus exposing their innate Hermiticity. Since birdtracks are a
graphical tool designed to make working with them more intuitive, it
is desirable to visually expose the inherent properties of the
birdtrack operators.

To illustrate how powerful these simplification rules are we show their effect on the Hermitian KS projector~\cite{Keppeler:2013yla} associated with the Young tableau
\begin{equation}
  \label{eq:MOLDAdvantageEx1}
\Phi :=
\begin{ytableau}
  1 & 2 & 4 & 7 \\
  3 & 6 \\
  5 & 8 \\
  9
\end{ytableau}
\ .
\end{equation}
The recursive KS algorithm leads to Hermitian Young projection
operator $P_{\Phi}$, with $\bar{P}_{\Phi}$ being of impressive length: It
contains 127 sets of symmetrizers and antisymmetrizers and its
Hermiticity is not visually apparent, see Figure~\ref{fig:MOLDAdvantage}. The cancellation rules achieve a
tremendous simplification: the result contains only 13
sets. Furthermore, multiple applications of the propagation rules can be used to
translate this into an explicitly symmetric form.
In fact, the shorter, explicitly symmetric result can be constructed
directly, without first constructing the KS-operator and then applying
the cancellation rules; we provide the tools to do so in a separate
paper~\cite{Alcock-Zeilinger:2016sxc}. The algorithm
allowing us to do this is the
MOLD-algorithm~\cite{Alcock-Zeilinger:2016sxc}. When
used to construct the Hermitean Young projection operators in
\emph{Mathematica}, the reduction in computational cost is impressive:
On a modern laptop, the final result shown in
Fig.~\ref{fig:MOLDAdvantage} is obtained approximately $18600$ times
faster than the KS-equivalent even before simplification rules are
implemented to shorten the result. Thus, the MOLD construction offers
a significant improvement over the KS construction. The proof of the
MOLD algorithm relies heavily on the manipulation rules laid out in
this paper.

Augmenting the Hermitean Young projection operators with transition
operators yields an alternative basis of the algebra of invariants of
$\SUN$ over $\Pow{m}$~\cite{Alcock-Zeilinger:2016cva}. The
construction algorithm of the transition operators again is built upon
the simplification rules presented in this paper. With a basis for
$\API{\SUN,\Pow{m}}$ consisting of projection and transition
operators, one can construct a \emph{mutually orthogonal, complete}
basis for the singlet states of $\SUN$ over $\MixedPow{m}{m}$ (and
more generally over
$\MixedPow{m}{n}$)~\cite{AlcockZeilinger2016Singlets}. These singlets
are directly applicable to QCD, as they are needed to form Wilson line
correlators used in the JIMWLK-framework~\cite{Marquet:2010cf} as
well as a modern treatment of jet-evolution equations (see
e.g.~\cite{Weigert:2003mm}), the infrared structure of QCD in form of
gluon exchange webs (see e.g.~\cite{Falcioni:2014pka}), GPD's and
TMD's (see e.g.~\cite{Bomhof:2006dp}). The references here only serve
to mark a specific reference point we find intriguing and are by no
means exhaustive: In fact, almost every branch of QCD in which the
factorization Theorems apply makes use of Wilson line correlators and
any attempt at completeness would be futile.

Besides their physics applications (which is the most appealing
quality to the authors of this paper), birdtracks are also immensely
useful in the study of the representation theory of semi-simple Lie
groups, as is exhibited in \cite{Cvitanovic:2008zz, Keppeler:2012ih,
  Keppeler:2013yla, Alcock-Zeilinger:2016sxc}. With the
recent interest in Hermitian Young projection operators, birdtracks
thus promise interesting further developement in this branch of
mathematics. It is hoped that the simplification rules given here
encourage the use of birdtracks as a viable tool for calculation.

\paragraph{Acknowledgements:} H.W. is supported by South Africa's
National Research Foundation under CPRR grant nr 90509. J.A-Z. was
supported (in sequence) by the postgraduate funding office of the
University of Cape Town (2014), the National Research Foundation (2015) and the
Science Faculty PhD Fellowship of the University of Cape Town (2016).

 \bibliographystyle{utphys}
 \bibliography{GroupTheory,PaperLibrary,BookLibrary}

\end{document}